\documentclass[11pt,a4paper]{article}
\usepackage{jheppub}
\usepackage{bm}
\usepackage{graphicx}
\usepackage{slashed}

\newcommand{\vek}[1]{\bm{#1}}
\newcommand{\fv}[1]{{#1}}
\newcommand{\imag}{\mathrm{i}}
\newcommand{\unit}[1]{\text{ #1}}
\newcommand{\dd}{\ensuremath{{\rm d}}}
\newcommand{\ckm}{{V}}
\newcommand{\openone}{\leavevmode\hbox{\small1\normalsize\kern-.33em1}}
\DeclareMathOperator{\im}{Im}
\DeclareMathOperator{\tr}{tr}
\DeclareMathOperator{\Tr}{Tr}

\title{Computing the temperature dependence of effective CP violation in the standard model}

\author[a,b]{Tom\'a\v{s} Brauner,}
\author[a]{Olli Taanila,}
\author[c]{Anders Tranberg}
\author[a]{and Aleksi Vuorinen}

\affiliation[a]{Faculty of Physics, University of Bielefeld, Bielefeld, Germany}
\affiliation[b]{Department of Theoretical Physics, Nuclear Physics Institute ASCR, \v{R}e\v{z}, Czech Republic}
\affiliation[c]{Niels Bohr International Academy, Niels Bohr Institute and Discovery Center, Copenhagen, Denmark}

\emailAdd{tbrauner@physik.uni-bielefeld.de}
\emailAdd{olli.taanila@iki.fi}
\emailAdd{anders.tranberg@nbi.dk}
\emailAdd{vuorinen@physik.uni-bielefeld.de}

\abstract{CP violation in the standard model originates from the Cabibbo-Kobayashi-Maskawa mixing matrix. Upon integrating all fermions out of the theory, its effects are captured by a series of effective nonrenormalizable operators for the bosonic gauge and Higgs fields. We compute the CP-violating part of the effective action to the leading nontrivial, sixth order in the covariant gradient expansion as a function of temperature. In the limit of zero temperature, our result addresses the discrepancy between two independent calculations existing in the literature~\cite{schmidt6,salcedo6}. We find that CP violation in the standard model is strongly suppressed at high temperature, but that at $T\lesssim1\unit{GeV}$ it may be relevant for certain scenarios of baryogenesis. We also identify a selected class of operators at the next, eighth order and discuss the convergence of the covariant gradient expansion.}

\keywords{CP violation, Standard Model, Thermal Field Theory}

\begin{document}

\maketitle

\section{Introduction}
\label{sec:intro}

The combination of the discrete parity and charge conjugation transformations (CP) is a symmetry of the gauge interactions of the standard model. However, a complex phase of the mixing matrix appearing in the fermion-scalar interactions breaks this symmetry. Combined with the breaking of baryon number conservation through the chiral anomaly~\cite{hooft}, this opens up the possibility of explaining the observed asymmetry between matter and antimatter in the universe, using only known electroweak-scale physics. This scenario is commonly referred to as electroweak baryogenesis~\cite{ewbg}. 

The dynamical processes leading to permanent change of baryon and lepton numbers in the standard model are intrinsically nonperturbative. As a consequence, we cannot expect to compute electroweak baryon number creation rates within a purely perturbative scheme. One approach to deal with this problem is to evolve the entire system numerically, for instance on a spacetime lattice. The advantage of this is that the out-of-equilibrium dynamics would be correctly treated, but a serious complication arises from the fact that the fermions need to obey quantum dynamics. This is numerically very challenging, as in addition to the usual issues associated with lattice fermions one needs to evolve each quantum fermion mode separately in time~\cite{aarts,PF1,PF2}. Fortunately, the gauge dynamics responsible for electroweak baryon number violation is very well understood numerically, both in and out of equilibrium, and can be convincingly described by classical dynamics or stochastic equations. The 
reason for this is that a change in baryon number is an infrared process, mediated by ``spatially large'' gauge field configurations like sphalerons. 

For the reasons listed above, it is natural to regard CP violation as a small, perturbative bias originating from the back-reaction of quantum fermions, living in a classical, nonperturbative gauge field background. Consequently, one may integrate out the quantum fermions in the path-integral formulation of the standard model, creating a purely bosonic effective theory of gauge and scalar fields. The effects of CP violation are then captured in a series of effective, nonrenormalizable operators. These can in turn be computed using a perturbative (or otherwise appropriate) expansion, their form and coefficients becoming functions of the fermion masses and mixing angles.

The benchmark calculation of the effective CP violation in the standard model was performed by Shaposhnikov~\cite{shaposhnikov}. Rather than integrating out fermions, his aim was to calculate the coefficient $\delta_\text{CP}$ of the leading CP-violating operator, defined in eq.~\eqref{FFtilde}, in perturbation theory in terms of diagrams with both bosonic and fermionic internal lines. 
The conclusion was that at asymptotically high temperatures $\delta_\text{CP}\simeq J\Delta/T^{12}$, while at zero temperature $\delta_\text{CP}\simeq J\Delta/v^{12}$, where $v\approx246\unit{GeV}$ is the Higgs field vacuum expectation value and $\Delta$ denotes a specific combination of fermion masses,
\begin{equation}
\Delta\equiv(m_u^2-m_c^2)(m_c^2-m_t^2)(m_t^2-m_u^2)(m_d^2-m_s^2)(m_s^2-m_b^2)(m_b^2-m_d^2)
\label{Delta}
\end{equation}
for quarks and analogously for leptons. In addition, $J$ stands here for the Jarlskog invariant, defined in terms of the $3\times 3$ fermion mixing matrix $\ckm$ (repeated indices not to be summed)
\begin{equation}
\im(\ckm_{ij}\ckm^{-1}_{jk}\ckm_{kl}\ckm^{-1}_{li})\equiv J\epsilon_{ik}\epsilon_{jl},
\label{jarlskog}
\end{equation}
where $\epsilon_{ij}\equiv\sum_k\epsilon_{ijk}$. In the quark sector, $\ckm$ represents the Cabibbo-Kobayashi-Maskawa (CKM) matrix, and $J\approx2.96\times 10^{-5}$~\cite{PDG}. Inserting the quark masses according to ref.~\cite{PDG}, $m_u=2.3\text{ MeV}$, $m_d=4.8\unit{MeV}$, $m_c=1.275\unit{GeV}$, $m_s=95\unit{MeV}$, $m_t=173.5\unit{GeV}$, $m_b=4.18\unit{GeV}$, one obtains $\delta_\text{CP}\simeq10^{-19}$ and $10^{-24}$ for $T=100\unit{GeV}$ and $T=0$, respectively. In combination with later work~\cite{Farrar1,Farrar2,Gavela1,Gavela2}, this simple estimate is the origin of the common lore that standard model CP violation is not sufficient to explain the observed baryon asymmetry, usually phrased as the net baryon-to-photon number ratio
\begin{equation}
\frac{n_\text{B}}{n_\gamma}\approx6\times10^{-10}.
\label{obsasym}
\end{equation}

Some time ago, Smit~\cite{smit4} demonstrated that the situation is in fact somewhat more complicated in that at least at zero temperature, integrating out the fermions does not lead to CP-violating operators suppressed by powers of the small Yukawa couplings, but that instead $\delta_{\text{CP}}\simeq J$. The first attempts to explicitly evaluate the leading CP-violating operators and their couplings, however, resulted in controversy: the operators found by Hernandez et al.~\cite{schmidt6} break P and conserve C, while all operators reported by Garc\'\i a-Recio and Salcedo~\cite{salcedo6} conserve P and break C.

While the non-suppression of effective CP violation at zero temperature is highly encouraging for baryogenesis, it is important to recall that realistic scenarios do not take place at zero temperature. ``Hot'' electroweak baryogenesis typically operates at temperatures around $100\unit{GeV}$~\cite{ewbg,HotBG}, while ``cold'' electroweak baryogenesis in an infrared-heavy out-of-equilibrium environment is assumed to take place at temperatures below roughly $40\unit{GeV}$~\cite{cewbag1,cewbag2,cewbag3,cewbag4}. (The temperature prior to electroweak symmetry breaking may be arbitrarily low, if emerging directly from an inflationary stage.) Hence, in the absence of a method to perform a fully out-of-equilibrium computation, knowledge of the (equilibrium) temperature dependence of effective CP violation in the bosonic sector of the standard model would clearly be desirable. It will allow us to connect directly to the equilibrium result of ref.~\cite{shaposhnikov}, but also approximately to the true out-of-equilibrium state, to the extent that the fermions can be described in terms of a smooth quasi-particle distribution function with an effective temperature. To this end, our present work is aimed at generalizing the calculations of refs.~\cite{smit4,schmidt6,salcedo6} to nonzero temperature and discussing their implications for the cold electroweak baryogenesis scenario. Some of the results derived here were presented already in ref.~\cite{PRL}.

The outline of our paper is as follows. In order not to obscure the physical picture with unnecessary technical details, we begin by providing a qualitative sketch of the calculation in section~\ref{sec:flow}. Section~\ref{sec:results} on the other hand presents our results in detail; in particular, section~\ref{sec:order6} reviews the temperature dependence of the leading CP-violating operators as reported in our previous paper~\cite{PRL}, while section~\ref{sec:order8} provides a partial list of subleading operators at zero temperature, and section~\ref{sec:leptons} investigates the effect of an additional source of CP violation in the lepton sector. The implications of our findings for baryogenesis are finally discussed in section~\ref{sec:appl}, while the technical details of our computation are relegated to several appendices. For the benefit of the reader, we also briefly summarize our notation and conventions in appendix~\ref{sec:notationSM}.

\section{Flow of the argument}
\label{sec:flow}

As explained above, our aim is to integrate out the fermions from the full standard model in order to derive an effective action for the gauge and Higgs fields, and finally to identify the CP-violating part thereof. As the effective action cannot be determined in a closed form for arbitrary spacetime-dependent background fields, we will resort to a covariant gradient expansion in its evaluation. For the spinodal transition involved in cold electroweak baryogenesis, simple estimates presented in section~\ref{sec:order8} will put very stringent bounds on how fast such a transition needs to be in order for the gradient expansion to be valid.

\subsection{Gradient expansion of the effective action}

Let us start by considering a generic Euclidean field theory of chiral fermions, denoted collectively as $\psi(\fv x)\equiv(\psi_{\text L}(\fv x),\psi_{\text R}(\fv x))$, coupled to external gauge as well as scalar fields. The partition function of such a theory can be expressed as a functional integral over the fermionic fields,
\begin{equation}
Z=\int[\dd\psi][\dd\bar\psi]\exp\left[-\int_{\fv x}\bar\psi(\fv x)\vek D(\fv x)\psi(\fv x)\right]\equiv e^{-\Gamma},
\end{equation}
where the effective action $\Gamma$ is formally given in terms of the Dirac operator $\vek D(\fv x)$ as $\Gamma=-\Tr\log\vek D$. However, as shown in refs.~\cite{salcedo1,salcedo2}, its evaluation is subtle due to renormalization ambiguities and the contribution from the chiral anomaly. To this end, we write the Dirac operator in the chiral (left-right) basis in a block form,
\begin{equation}
\vek D=\begin{pmatrix}
\slashed{D}_\text{L} & m_\text{LR}\\
m_\text{RL} & \slashed{D}_\text{R}
\end{pmatrix},
\label{eq:origD}
\end{equation}
and define the object $K\equiv m_\text{LR}m_\text{RL}-\slashed{D}_\text{L}m_\text{RL}^{-1}\slashed{D}_\text{R}m_\text{RL}$. The effective action is then given by the sum of a normal and an abnormal parity component, $\Gamma^+$ and $\Gamma^-$, expressed in terms of $K$ as
\begin{equation}
\Gamma^+=-\frac12\mathrm{Re}\Tr(\log K),\qquad
\Gamma^-=-\frac\imag2\im\Tr(\gamma_5\log K)+\Gamma_{\text{gWZW}},
\label{effactiongeneral}
\end{equation}
where $\Gamma_{\text{gWZW}}$ is the anomalous Wess-Zumino-Witten term~\cite{salcedo1,salcedo2}. 

The dominant source of CP violation in the standard model is the CKM matrix in the quark sector, to which we will restrict our attention from now on. (The effect of additional CP violation originating from the lepton sector will be discussed in section~\ref{sec:leptons}.) In the block form indicated in eq.~\eqref{eq:origD}, the quark Dirac operator reads
\begin{equation}
\vek D=\left(\begin{array}{cc|cc}
\slashed{D}_u+\slashed{Z}+\slashed{G} & \slashed{W}^+ & \frac\phi vM_u & 0 \\
\slashed{W}^- & \slashed{D}_d-\slashed{Z}+\slashed{G} & 0 & \frac\phi vM_d\\
\hline
\frac\phi vM_u^\dagger & 0 & \slashed{D}_u+\slashed{G} & 0\\
0 & \frac\phi vM_d^\dagger & 0 & \slashed{D}_d+\slashed{G}
\end{array}\right),
\label{eq:SMmat}
\end{equation} 
cf.~eq.~\eqref{SMdirac}. Here, $Z_\mu$, $W^\pm_\mu$ and $G_\mu$ are the weak intermediate boson and gluon fields, respectively, $\phi$ is the Higgs field, and $M_{u,d}$ stand for the complex mass matrices of $u$-type and $d$-type quarks before diagonalization. Further notation, including explicit expressions for the covariant derivatives $D^{u,d}_\mu$, is summarized in appendix~\ref{sec:notationSM}. Throughout this paper, we use the unitary gauge, in which only a single physical scalar field $\phi$ appears.

Using the explicit expression for the quark operator~\eqref{eq:SMmat}, we obtain for the operator~$K$
\begin{equation}
K=\begin{pmatrix}
(\phi^2/v^2)M_uM_u^\dag-(\slashed D_u+\slashed Z)(\slashed D_u+\slashed\varphi) & -\slashed W^+(\slashed D_d+\slashed\varphi)\\
-\slashed W^-(\slashed D_u+\slashed\varphi) & (\phi^2/v^2)M_dM_d^\dag-(\slashed D_d-\slashed Z)(\slashed D_d+\slashed\varphi)
\end{pmatrix},
\label{Kop}
\end{equation}
with $\varphi_\mu\equiv\phi^{-1}[\partial_\mu,\phi]$.\footnote{Recall that at this stage of the computation, we are dealing with (pseudo-)differential operators. All fields act as multiplicative operators and the commutator $[\partial_\mu,\phi]$ thus corresponds to a multiplication by a derivative of the Higgs field.} The gluon fields have now been omitted for the sake of simplicity, as they only contribute at the subleading, eighth order in the gradient expansion~\cite{salcedo8}. One nevertheless has to bear in mind that the trace in eq.~\eqref{effactiongeneral} still involves the color space, thus giving rise to the trivial prefactor $N_{\text c}$, the number of colors.

In the next step one expands the logarithm of $K$ in powers of its off-diagonal part, or equivalently, in powers of the $W^\pm_\mu$ fields. Following ref.~\cite{salcedo6}, let us introduce a shorthand notation for the inverse of the diagonal components of $K$,
\begin{equation}
\tilde N_{u,d}^{-1}\equiv\frac{\phi^2}{v^2}M_{u,d}M_{u,d}^\dagger-(\slashed D_{u,d}\pm\slashed Z)(\slashed D_{u,d}+\slashed\varphi),\qquad
N_{u,d}^{-1}\equiv\frac{\phi^2}{v^2}M_{u,d}M_{u,d}^\dagger+\fv p^2.
\label{eq:master_prop}
\end{equation}
One obtains a compact expression $\langle\log K\rangle=\sum_{n=0}^\infty\langle\log K\rangle_{2n}$, where $\langle\mathcal{O}\rangle$ stands for $\Tr\mathcal{O}$ or $\Tr(\gamma_5\mathcal{O})$ in $\Gamma^+$ and $\Gamma^-$, respectively, and
\begin{equation}
\langle\log K\rangle_{2n}\equiv -\frac1n\left\langle\bigl[\tilde N_u\slashed W^+(\slashed D_d+\slashed\varphi)\tilde N_d\slashed W^-(\slashed D_u+\slashed\varphi)\bigr]^n\right\rangle.
\label{master_covariant}
\end{equation}
The trace is evaluated with the help of the method of covariant symbols, explained in detail in appendix~\ref{sec:symbols}. In the covariant gradient expansion, the result is further expanded in powers of the covariant derivatives and the $Z_\mu,\varphi_\mu$ fields. All the contributions are then classified as $\langle\log K\rangle_{2n+m}$ where, as above, $2n$ is the number of $W^\pm_\mu$ fields and $m$ counts the covariant derivatives, $Z_\mu$ and $\varphi_\mu$ fields. The order of a given operator in the gradient expansion is $2n+m$.

In order to have any CP violation, at least four $W^\pm_\mu$ fields are needed. This is because the simplest CP-violating observable built from the CKM matrix is the Jarlskog invariant, eq.~\eqref{jarlskog}, which is composed of a product of four CKM matrix elements. Therefore, CP violation can only appear at order four or higher in the gradient expansion. However, as was shown by Smit~\cite{smit4}, order four actually gives no CP violation, a result which we verified to hold also at nonzero temperature. Since the chiral anomaly only contributes at this order in the gradient expansion, it can be discarded as long as only the CP-violating part of the effective action is desired~\cite{salcedo6}. We can thus use eq.~\eqref{master_covariant} to identify all effective bosonic CP-violating operators of the standard model.

At zero temperature (and in an even-dimensional spacetime), only operators of even orders in gradients appear in the action as a consequence of Lorentz invariance. Nonzero temperature can in principle lead to operators of the 4+1 type, but we checked by an explicit computation that they do not contribute to CP violation in the standard model. The leading order at which CP violation appears is then order six. In fact, there are no contributions of the $2n$+0 type for any $n$, as demonstrated by Garc\'\i a-Recio and Salcedo~\cite{salcedo6} and verified by us also at nonzero temperature. Therefore, all contributions at the leading nontrivial order must be of the type 4+2; they are investigated in detail in section~\ref{sec:order6}. In order to assess the convergence of the gradient expansion, we evaluated a selected class of operators at the next, eighth order at zero temperature. There is no CP violation coming from operators of the type 8+0, which leaves us with two possibilities: 6+2 and 4+4 type operators. In section~\ref{sec:order8} and the associated appendix~\ref{sec:jarlskog6}, we derive the full CP-violating action in the 6+2 sector. We leave aside the evaluation of the 4+4 operators since the immense number of possible combinations of Lorentz indices and fields makes it impossible to even list all contributing operators in a paper. We merely note that the first CP-violating and P-violating operators calculated in ref.~\cite{salcedo8} are of the 4+4 type.
 
\subsection{Remarks on the gradient expansion at nonzero temperature}
\label{sec:symboldisc}

While the above discussion lays down a concrete path both at zero and nonzero temperature, there is one additional subtlety in the calculation that needs to be addressed. It is namely a well-known fact that the gradient expansion breaks down at nonzero temperature. This may be naively understood as a consequence of the periodic boundary condition imposed on the quantum fields in the Matsubara formalism; for any given bosonic field $\phi$ as a function of the imaginary time $\tau$, this reads $\phi(\tau=1/T)=\phi(\tau=0)$, where $T$ is the temperature. Expanding $\phi(\tau=1/T)-\phi(\tau=0)$ in Taylor series around $\tau=0$, we obtain a series of terms of successively higher order in the gradients, which only produce zero upon the full resummation. Truncating the expansion at any finite order gives spurious contributions that, apart from the periodicity itself, may break other symmetries of the effective action. The fully gauge invariant effective action is necessarily nonlocal; the compactification of the time dimension results in the appearance of the Polyakov loop operators~\cite{Megias,salcedoT}.  

When computing the bosonic effective action of the standard model at nonzero temperature, the above problem shows up in those terms containing temporal derivatives of the various fields. While the proper way to deal with them would be to follow the strategy of ref.~\cite{salcedoT}, we took a simpler approach, avoiding the necessity to cope with the unphysical contributions that possibly break the symmetries of the effective action. Concretely, we performed on purpose the gradient expansion also in the forbidden temporal derivatives and restricted our attention to the Lorentz-invariant part of the result, which does not suffer from the above problem. It can be defined unambiguously by discarding all operators containing temporal indices (whether in derivatives or gauge fields), and then replacing the remaining spatial indices with Lorentz spacetime ones. This should be kept in mind when inspecting the results of section~\ref{sec:order6}.

\subsection{Momentum integration}
 
The evaluation of the effective action using the method of (covariant) symbols leads to momentum integrals of the type 
\begin{equation}
\int_p(\vek p^2)^k(p_0^2)^l\tr\bigl(N_u^{r_1}N_d^{t_1}\dotsb N_u^{r_n}N_d^{t_n}\bigr),
\label{Iintaux}
\end{equation}
where the integration symbol represents either an integral over all spacetime components of momentum (at zero temperature) or an integral over spatial components together with a Matsubara sum over the temporal component (at nonzero temperature). For full details of our notation, see appendix~\ref{sec:notationSM}. The dependence of the result on the Higgs field $\phi$ is trivial, as can be seen by rescaling the spatial momentum $\vek p$ by $\phi/v$ and \emph{redefining} the temperature to $T_\text{eff}\equiv Tv/\phi$. We thus see that the above momentum integral can be written in the form $(\phi/v)^\lambda I^{k,l}_{r_1,t_1,\dotsc,r_n,t_n}$, where $\lambda\equiv d+2(k+l)-2\sum_{i=1}^n(r_i+t_i)$, $d$ is the spacetime dimension, and
\begin{equation}
\begin{split}
I^{k,l}_{r_1,t_1,\dotsc,r_n,t_n}&\equiv\int_p(\vek p^2)^k(p_0^2)^l\\
&\times\tr\frac1{(\fv p^2+M_uM_u^\dagger)^{r_1}(\fv p^2+M_dM_d^\dagger)^{t_1}\dotsb(\fv p^2+M_uM_u^\dagger)^{r_n}(\fv p^2+M_dM_d^\dagger)^{t_n}}.
\end{split}
\label{IintT}
\end{equation}
The temporal component of momentum in this integral is determined by $T_\text{eff}$ rather than $T$ itself; this should be kept in mind throughout the detailed derivation of some of the thermal integrals in appendix~\ref{sec:integrals}. An analogous argument applies at zero temperature, the only difference being that the whole four-momentum $\fv p$ must be rescaled by $\phi/v$. In addition, Lorentz invariance then ensures that the integrals $I^{k,l}_{r_1,t_1,\dotsc,r_n,t_n}$ appear only in specific combinations, discussed at length in section~5 of ref.~\cite{salcedo6},
\begin{align}
\notag
I^{k}_{r_1,t_1,\dotsc,r_n,t_n}&\equiv\int_p(\fv p^2)^k\tr\frac1{(\fv p^2+M_uM_u^\dagger)^{r_1}(\fv p^2+M_dM_d^\dagger)^{t_1}\dotsb(\fv p^2+M_uM_u^\dagger)^{r_n}(\fv p^2+M_dM_d^\dagger)^{t_n}}\\
&=\sum_{l=0}^k\binom klI^{l,k-l}_{r_1,t_1,\dotsc,r_n,t_n}.
\label{Iint}
\end{align}
In the effective action, this is accompanied by the factor $(\phi/v)^\lambda$ where now $\lambda\equiv d+2k-2\sum_{i=1}^n(r_i+t_i)$.

As the CP transformation is equivalent to complex conjugation of the mass matrices, $M_{u,d}$, the CP-violating part of the effective action is proportional to the imaginary part of the momentum integrals, $\hat I^{k,l}_{r_1,t_1,\dotsc,r_n,t_n}\equiv\imag\im I^{k,l}_{r_1,t_1,\dotsc,r_n,t_n}$ and analogously for $\hat I^{k}_{r_1,t_1,\dotsc,r_n,t_n}$. At zero temperature, the leading-order CP-violating part of the effective action (belonging to the 4+2 sector) is proportional to the single integral $\hat I^3_{1,1,2,2}$. The action in the 6+2 sector turns out to be proportional to a single integral as well, this time $\hat I^4_{1,1,1,1,2,2}$; some details of its calculation are presented in appendix~\ref{sec:jarlskog6}. The evaluation of the temperature dependence of the leading-order result on the other hand requires the determination of the integrals $\hat I^{k,3-k}_{1,1,2,2}$ as well as $\hat I^{k,4-k}_{1,1,2,3}$ and $\hat I^{k,4-k}_{1,1,3,2}$, with non-negative integer superscripts. Their calculation is detailed in appendix~\ref{sec:integrals}.

\section{Results}
\label{sec:results}

In this section, we list all of the results we have obtained at different orders of the gradient expansion both at zero and nonzero temperature. In practice, the computations were performed using the Mathematica package Feyncalc~\cite{Feyncalc}. Our code uses the method of covariant symbols to evaluate the trace in eq.~\eqref{master_covariant}, taking as input the expansion coefficients from eqs.~\eqref{SMcovsymb} and~\eqref{Ucoeff}. The resulting momentum integrals were simplified manually and subsequently evaluated numerically. While the general derivations presented in this paper are formulated and valid in a Euclidean spacetime of an arbitrary dimension, the results presented in this section are specific to the physical case of four dimensions.

It is worthwhile to remark that in order to cast the effective action into the canonical form presented below, integration by parts (in coordinate space) was sometimes necessary to avoid, for instance, the appearance of operators with all derivatives acting on a single field. When doing so, we tacitly assumed that no surface terms arise from such integration. This is equivalent to demanding that the fields are topologically trivial, which is consistent with the fact that we are performing a gradient expansion. On a more technical note, when integrating by parts, one has to pay attention to the fact that the mass matrices in the propagators~\eqref{eq:master_prop} contain a factor of the Higgs field $\phi$. At zero temperature, this can be easily gotten rid of by rescaling the momentum variable prior to the integration. This is no longer possible at nonzero temperature, and we carried out the integration by parts before the step leading from eq.~\eqref{Iintaux} to eq.~\eqref{IintT}. This results in particular in the appearance of momentum integrals of the type~\eqref{I1123}.

\subsection{Leading order}
\label{sec:order6}

At zero temperature, the CP-violating part of the effective action was evaluated to its leading, sixth order in refs.~\cite{schmidt6,salcedo6}. The main goal of our preceding paper~\cite{PRL} was to resolve the discrepancy between these two calculations and to generalize the results to finite temperature. As observed in the previous section, all relevant CP-violating operators at this order are of the 4+2 type, and hence the resulting (Euclidean) effective action takes the form
\begin{equation}
\Gamma_{\text{CP-odd}}^{\text{4+2}}=-\frac\imag2N_{\text c}JG_{\text{F}}\kappa_{\text{CP}}^{\text{4+2}}\int_{\fv x}\left(\frac v\phi\right)^2({\mathcal O}_0+{\mathcal O}_1+{\mathcal O}_2),
\label{action1}
\end{equation}
where $G_{\text{F}}=1/(\sqrt2v^2)$ is the Fermi coupling, the coefficient $\kappa_{\text{CP}}^{\text{4+2}}$ is defined as\footnote{We write the integration measure explicitly to emphasize that $\kappa_\text{CP}^\text{4+2}$ is defined at zero temperature; the temperature dependence of the CP violation effects originates through the coefficients $c_i$ introduced below. The coefficient $\kappa_\text{CP}^\text{4+2}$ is expressed in terms of a momentum integral of the type~\eqref{Iint} as $\imag JG_\text{F}\kappa^\text{4+2}_\text{CP}=\hat I^3_{1,1,2,2}$. Due to the hierarchy of the quark masses, it can be well approximated as $\kappa_\text{CP}^\text{4+2}\approx1/(16\pi^2G_\text{F}m_c^2)\approx334$.}
\begin{equation}
\kappa_{\text{CP}}^\text{4+2}\equiv\frac{\Delta}{G_{\text{F}}}\int\frac{\dd^4\fv p}{(2\pi)^4}\frac{(\fv p^2)^3}{\prod_{q=1}^6(\fv p^2+m_q^2)^2}\approx309,
\label{kappaCP}
\end{equation}
and the parameter $\Delta$ was introduced in eq.~\eqref{Delta}. 

The operators $\mathcal O_n$ appearing above are composed of the $W^\pm_\mu$, $Z_\mu$ and $\varphi_\mu$ fields; the subscript $n$ counts the number of the latter two. They can be further split into P-even and P-odd parts,  $\mathcal O_n=\mathcal O^+_n+\mathcal O^-_n$. At nonzero temperature, the effective action is no longer Lorentz-invariant due to the presence of the thermal bath, but can be formally written in a covariant way, if one introduces a timelike vector defining the rest frame of the thermal bath, $u_\mu\equiv\delta_{\mu0}$. We can then also divide each operator $\mathcal O_n$ into its Lorentz-invariant part, containing no $u_\mu$'s, and the rest. We have verified that the Lorentz-noninvariant part of the action vanishes in the limit of zero temperature, as it must. The Lorentz-invariant part is on the other hand explicitly given by ${\mathcal O}_0^-={\mathcal O}_1^-={\mathcal O}_2^-=0$ and
\begin{align}
\notag{\mathcal O}_0^+=&-\frac{c_1}{3}(W^+)^2W^-_{\mu\mu}W^-_{\nu\nu}+\frac{5c_2}{3}(W^+)^2W^-_{\mu\nu}W^-_{\mu\nu}-\frac{c_1}{3}(W^+)^2W^-_{\mu\nu}W^-_{\nu\mu}\\
\label{O0}&+\frac{4c_3}{3}W^+_\mu W^+_\nu W^-_{\mu\nu}W^-_{\alpha\alpha}-\frac{2c_1}{3}W^+_\mu W^+_\nu W^-_{\mu\alpha}W^-_{\nu\alpha}\\
\notag&-2c_4W^+_\mu W^+_\nu W^-_{\alpha\mu}W^-_{\alpha\nu}+\frac{4c_3}{3}W^+_\mu W^+_\nu W^-_{\mu\alpha}W^-_{\alpha\nu}-\text{c.c.} ,\\
\notag{\mathcal O}_1^+=&\frac83(Z_\mu+\varphi_\mu)\bigl[c_5(W^+)^2W^-_\mu W^-_{\nu\nu}-c_6(W^+)^2W^-_\nu W^-_{\mu\nu}-c_6(W^+)^2W^-_\nu W^-_{\nu\mu}\\
\label{O1}&-c_3(W^+\cdot W^-)W^+_\mu W^-_{\nu\nu}+c_7(W^+\cdot W^-)W^+_\nu W^-_{\mu\nu}+c_7W^+_\mu W^+_\nu W^-_\alpha W^-_{\alpha\nu}\\
\notag&-c_{12}(W^+\cdot W^-) W^+_\nu W^-_{\nu\mu} -c_{12} W^+_\mu W^+_\nu W^-_\alpha W^-_{\nu\alpha}+c_{13} W^-_\mu W^+_\nu W^+_\alpha W^-_{\nu\alpha}\bigr]-\text{c.c.} ,\\
\notag{\mathcal O}_2^+=&4(Z_\mu Z_\nu+\varphi_\mu\varphi_\nu)\bigl[c_8(W^+)^2W^-_\mu W^-_\nu-c_8(W^-)^2W^+_\mu W^+_\nu\bigr]\\
\label{O2}&-\frac{16}3(Z\cdot\varphi)\bigl[c_9(W^+\cdot W^-)^2-2c_6(W^+)^2(W^-)^2\bigr]+\frac43(Z_\mu\varphi_\nu+Z_\nu\varphi_\mu)\\
\notag&\times\bigl[c_{10}(W^+)^2W^-_\mu W^-_\nu+c_{10}(W^-)^2W^+_\mu W^+_\nu-2c_{11}(W^+\cdot W^-)(W^+_\mu W^-_\nu+W^+_\nu W^-_\mu)\bigr],
\end{align}
where ``$\text{c.c.}$'' stands for complex conjugation. The coefficients $c_i$ are complicated functions of the quark masses, temperature and the local Higgs field $\phi(\fv x)$, of which the latter two appear in the combination $T_{\text{eff}}=Tv/\phi$. The list of explicit expressions for the coefficients is lengthy and we thus provide it at the end of appendix~\ref{sec:integrals}, in eq.~\eqref{cs}. In the zero-temperature limit, the coefficients $c_1$--$c_{11}$ approach unity while $c_{12},c_{13}$ vanish, reducing our result to that of ref.~\cite{salcedo6}. We emphasize that all the operators written above are P-even and C-odd, generalizing the observation of ref.~\cite{salcedo6} to finite temperature.

\begin{figure}[t]
\begin{center}
\includegraphics[width=0.7\textwidth]{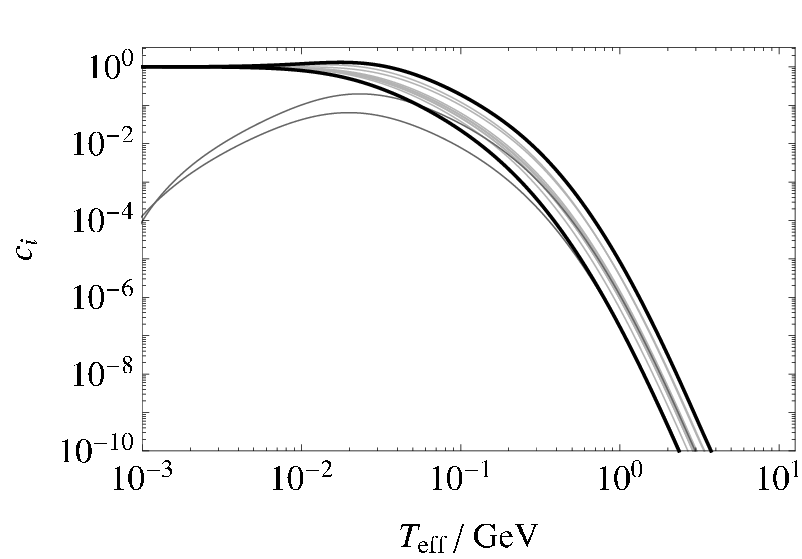}
\caption{The coefficients $c_1$--$c_{13}$ plotted as functions of the effective temperature $T_\text{eff}=Tv/\phi$. The thick lines correspond to the smallest and the largest of those $c_i$'s that approach one at zero temperature, that is, $c_1$ and $c_{10}$. This figure was previously published in ref.~\cite{PRL}.}
\label{fig1}
\end{center}
\end{figure}

\begin{figure}[t]
\begin{center}
\includegraphics[width=0.7\textwidth]{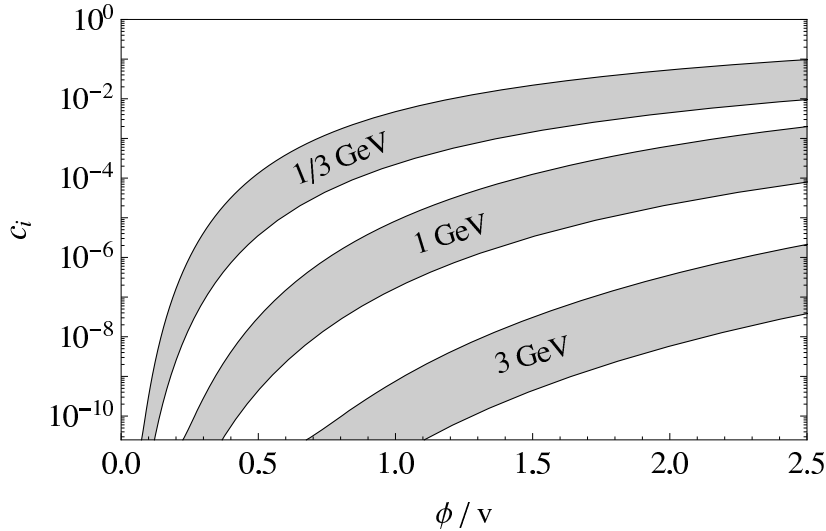}
\caption{The dependence of the $c_i$'s on the Higgs field $\phi$, plotted for three different values of the temperature, $T=1/3\unit{GeV}$, $1\unit{GeV}$ and $3\unit{GeV}$. In each case, the grey band is spanned by $c_1$ and $c_{10}$. This figure was previously published in ref.~\cite{PRL}.}
\label{fig2}
\end{center}
\end{figure}

The behavior of the coefficients $c_i$ as functions of $T_\text{eff}$ is displayed in figs.~\ref{fig1} and \ref{fig2}, reproduced from ref.~\cite{PRL}. Our main observation concerns the rate at which these functions fall as one progresses towards higher temperatures: while at $T_\text{eff}\lesssim100\unit{MeV}$ the coefficients display a relatively slow variation, at phenomenologically interesting higher temperatures they start to decrease very rapidly with $T_\text{eff}$. The largest coefficient at all temperatures is $c_{10}$, which reaches its maximal value of approximately $1.3$ at $T_\text{eff}\approx18\unit{MeV}$, while at $1\unit{GeV}$ it has already decreased down to the $10^{-5}$ level. At even higher temperatures, we observe a temperature dependence of the coefficients which is compatible with the estimate of ref.~\cite{shaposhnikov}.

As highlighted in figure~\ref{fig2}, the dependence of the coefficients $c_i$ on the parameter $T_\text{eff}$ has the impact that at any fixed nonzero temperature, their magnitude depends very strongly on the value of the Higgs field $\phi$. Small values of the Higgs field correspond to stronger suppression of CP violation, while those regions of space where $\phi$ is large witness an enhancement of CP violation. As discussed at some length in ref.~\cite{PRL}, this observation has the effect of improving the convergence of the gradient expansion, as it alleviates the problems numerical simulations of standard model dynamics face in those regions of space, where the Higgs field changes its sign.

\subsection{Next-to-leading order}
\label{sec:order8}

The motivation to investigate the next-to-leading-order terms in the gradient expansion is twofold. First, as observed in ref.~\cite{salcedo6} and confirmed by us, all CP-violating operators at the leading order conserve parity. This is a very nontrivial and unexpected result, which is at odds with the findings of ref.~\cite{schmidt6}. At present, we cannot offer a satisfactory explanation for why all P-odd contributions, though present in the intermediate stages of the calculation, should cancel in the final result. It is then important to try to detect the first P-odd operators in the gradient expansion. A selected class of such operators, of the type $(W^+W^-)^2Z^3D$, was indeed found by Salcedo at the eighth order~\cite{salcedo8}. His result, albeit incomplete, is certainly valuable, calling for both verification and generalization. A second, altogether separate, motivation for the next-to-leading-order calculation is that it allows us to examine the convergence of the gradient expansion. This is clearly a highly relevant issue for all practical applications of our results.

As observed in the previous section, the order-eight operators are either of the type 6+2 or 4+4, the former arising from the $n=3$ term of eq.~\eqref{master_covariant}. At zero temperature, the full result in the 6+2 sector reads
\begin{equation}
\Gamma_\text{CP-odd}^{\text{6+2}}=\frac4{15}N_\text{c}\hat I^4_{1,1,1,1,2,2}\int_{\fv x}\left(\frac v\phi\right)^4\left(\mathcal{O}_0+\mathcal{O}_1+\mathcal{O}_2\right),
\end{equation}
where again ${\mathcal O}_0^-={\mathcal O}_1^-={\mathcal O}_2^-=0$ and
\begin{align}
\notag\mathcal{O}_0^+=&(W^+)^2(W^+\cdot W^-)(W^-_{\mu\mu}W^-_{\nu\nu}-4W^-_{\mu\nu}W^-_{\mu\nu}+W^-_{\mu\nu}W^-_{\nu\mu})+2(W^+)^2W^+_\mu W^-_\nu\\
&\times(W^-_{\mu\nu}W^-_{\alpha\alpha}+W^-_{\nu\mu}W^-_{\alpha\alpha}+W^-_{\mu\alpha}W^-_{\nu\alpha}-4W^-_{\alpha\mu}W^-_{\alpha\nu}+W^-_{\mu\alpha}W^-_{\alpha\nu}+W^-_{\alpha\mu}W^-_{\nu\alpha})\\
\notag&+6(W^+\cdot W^-)W^+_\mu W^+_\nu\bigl(-W^-_{\mu\nu}W^-_{\alpha\alpha}+\tfrac13W^-_{\mu\alpha}W^-_{\nu\alpha}+2W^-_{\alpha\mu}W^-_{\alpha\nu}-W^-_{\mu\alpha}W^-_{\alpha\nu}\bigr)\\
\notag&+4W^+_\mu W^+_\nu W^+_\alpha W^-_\beta W^-_{\mu\nu}W^-_{\alpha\beta}-6W^+_\mu W^+_\nu W^+_\alpha W^-_\beta W^-_{\mu\nu}W^-_{\beta\alpha}-\text{c.c.},\\
\notag\mathcal{O}_1^+=&2(Z_\mu+\varphi_\mu)\bigl[-2(W^+)^2(W^-)^2W^+_\mu W^-_{\nu\nu}+3(W^+)^2(W^-)^2W^+_\nu W^-_{\mu\nu}\\
\notag&-2(W^+)^2(W^-)^2W^+_\nu W^-_{\nu\mu}-4(W^+)^2(W^+\cdot W^-)W^-_\mu W^-_{\nu\nu}\\
\notag&+6(W^+)^2(W^+\cdot W^-)W^-_\nu W^-_{\mu\nu}+(W^+)^2(W^+\cdot W^-)W^-_\nu W^-_{\nu\mu}\\
&+6(W^+\cdot W^-)^2W^+_\mu W^-_{\nu\nu}-9(W^+\cdot W^-)^2W^+_\nu W^-_{\mu\nu}+(W^+\cdot W^-)^2W^+_\nu W^-_{\nu\mu}\\
\notag&+(W^+)^2W^+_\mu W^-_\nu W^-_\alpha W^-_{\nu\alpha}+(W^+)^2W^-_\mu W^+_\nu W^-_\alpha W^-_{\nu\alpha}+(W^+)^2 W^-_\mu W^-_\nu W^+_\alpha W^-_{\nu\alpha}\\
\notag&+(W^-)^2W^+_\mu W^+_\nu W^+_\alpha W^-_{\nu\alpha}-3(W^+\cdot W^-)W^+_\mu W^+_\nu W^-_\alpha W^-_{\nu\alpha}\\
\notag&-3(W^+\cdot W^-)W^+_\mu W^-_\nu W^+_\alpha W^-_{\nu\alpha}+2(W^+\cdot W^-)W^-_\mu W^+_\nu W^+_\alpha W^-_{\nu\alpha}\bigr]-\text{c.c.},\\
\notag\mathcal{O}_2^+=&10(Z_\mu Z_\nu+\varphi_\mu\varphi_\nu)(W^+\cdot W^-)\bigl[(W^-)^2W^+_\mu W^+_\nu-(W^+)^2W^-_\mu W^-_\nu\bigr]\\
&+32(Z\cdot\varphi)(W^+\cdot W^-)\bigl[(W^+\cdot W^-)^2-(W^+)^2(W^-)^2\bigr]\\
\notag&+2(Z_\mu\varphi_\nu+Z_\nu\varphi_\mu)\bigl\{-(W^+)^2(W^+\cdot W^-)W^-_\mu W^-_\nu-(W^-)^2(W^+\cdot W^-)W^+_\mu W^+_\nu\\
\notag&+\bigl[4(W^+\cdot W^-)^2-3(W^+)^2(W^-)^2\bigr](W^+_\mu W^-_\nu+W^+_\nu W^-_\mu)\bigr\}.
\end{align}
Note that similarly to the type 4+2 operators, there are no P-odd contributions. The prefactor of the result is given by a new type of integral, the detailed derivation of which is performed in appendix~\ref{sec:jarlskog6}. Similarly to the prefactor in eq.~\eqref{action1}, it can be expressed in terms of a dimensionless parameter, $\kappa_\text{CP}^\text{6+2}$, defined as $(4/15)\hat I^4_{1,1,1,1,2,2}\equiv\imag JG_\text{F}^2\kappa_\text{CP}^\text{6+2}$. We find 
\begin{equation}
\kappa_\text{CP}^\text{6+2}=\frac4{15}\frac\Delta{G_\text{F}^2}\int\frac{\dd^4\fv p}{(2\pi)^4}\frac{(\fv p^2)^4}{\prod_{q=1}^6(\fv p^2+m_q^2)^2}\sum_{i,j}\frac{|\ckm_{ij}|^2}{(\fv p^2+m_{u,i}^2)(\fv p^2+m_{d,j}^2)}\approx2.65\times10^9,
\label{kappa62}
\end{equation}
see eq.~\eqref{integral62}. Interestingly, this integral is strongly infrared sensitive. It would diverge if we naively tried to set the $u$- and $d$-quark masses to zero, and in fact, $99\%$ of its value comes from the $i=j=1$ term in the sum. As in the case of $\kappa_\text{CP}^\text{4+2}$, a very good approximation is obtained by factorizing out all the heavy quark masses, producing
\begin{equation}
\kappa_\text{CP}^\text{6+2}\approx\frac{|\ckm_{ud}|^2}{30\pi^2G_\text{F}^2m_c^2m_s^2}\left(\log\frac{m_s}{\bar m}-\frac{197}{120}\right)\approx2.64\times10^9,
\end{equation}
where $\bar m\equiv(m_u+m_d)/2$.

We have not determined the exhaustive list of CP-violating operators of the 4+4 type. The reason for not having completed this calculation is twofold. First, the result is plagued with ambiguities due to the existence of relations among different operators arising from integration by parts. (The approach used in ref.~\cite{salcedo8} seems more suitable for this purpose as it derives the effective action from the conserved current and thus avoids such ambiguities by construction.) This is, however, merely a technical obstacle, which can be overcome with some effort. The reason we have not attempted to do so is that the full CP-violating action in the 4+4 sector contains several thousand independent operators. While the coefficients of some of them probably vanish due to accidental cancellations, the result would still be of little practical use for subsequent numerical simulations, and would almost certainly not be possible to publish in a journal article.

Finally, let us remark that at order eight, the gluon fields start to contribute to the CP-violating operators. All terms including gluons have been identified by Salcedo~\cite{salcedo8}, and we have confirmed his result. The possible form of such operators is strongly constrained by symmetries: two factors of the gluon field strength tensor are needed in order to construct a color singlet, which together with the four $W^\pm_\mu$ fields necessary for CP violation make for an order-eight operator; no $Z_\mu$ or $\varphi_\mu$ fields can thus appear. The final result for the gluonic part of the Euclidean CP-violating effective action reads
\begin{equation}
\Gamma^{\text{4+gluon}}_{\text{CP-odd}}=\frac83\hat I^2_{1,1,2,2}\int_{\fv x}\left(\frac v\phi\right)^4\bigl[(W^+)^2W^-_\mu W^-_\nu-(W^-)^2W^+_\mu W^+_\nu\bigr]\tr(G_{\alpha\mu}G_{\alpha\nu}),
\end{equation}
where $G_{\mu\nu}\equiv[\partial_\mu+G_\mu,\partial_\nu+G_\nu]$ is the gluon field strength tensor and the trace is performed only over the color space. Our result is in agreement with that of ref.~\cite{salcedo8}.

\begin{figure}[t]
\begin{center}
\includegraphics[width=0.7\textwidth]{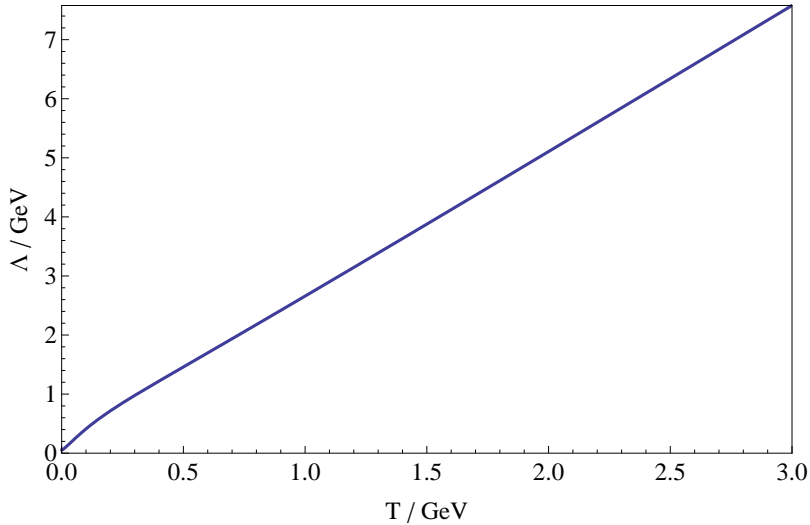}
\caption{The behavior of the scale $\Lambda$, defined in eq.~\eqref{lambdadef}, as a function of the effective temperature $T_\text{eff}$.}
\label{Lambda}
\end{center}
\end{figure}

In order to investigate the convergence of the gradient expansion both at zero and nonzero temperature, we use the result for the effective coupling in the 6+2 sector~\eqref{kappa62}, and compare it naively to $\kappa_\text{CP}^\text{4+2}$, eq.~\eqref{kappaCP}. Generalizing both expressions to finite temperature by simply replacing the four-dimensional momentum integrals by sum-integrals, we define a parameter with the dimension of energy, $\Lambda$, through
\begin{equation}
\Lambda^2\equiv\frac{\kappa_\text{CP}^\text{4+2}}{2G_\text{F} \kappa_\text{CP}^\text{6+2}}.
\label{lambdadef}
\end{equation}
We expect the gradient expansion to converge if the characteristic magnitude of the gauge fields is smaller than this scale. The temperature dependence of the parameter $\Lambda$ is displayed in figure~\ref{Lambda}. We observe a rather steep increase of $\Lambda$ as a function of temperature: while at zero temperature, it acquires the relatively low value $53\unit{MeV}$, at $T_\text{eff}=1\unit{GeV}$ it has already reached the value $2.6\unit{GeV}$. This is due to the fact that $\kappa_\text{CP}^\text{6+2}$ is suppressed even more strongly than $\kappa_\text{CP}^\text{4+2}$ at high temperature. Thus, we can conclude that nonzero temperature tends to improve the convergence of the gradient expansion.

The definition of the scale $\Lambda$ needs to be complemented by an evaluation of the magnitude of the gauge fields in the order-six and order-eight operators. Ideally, we would want to compute numerically the relative size of these operators during a slow spinodal transition. Such a calculation is beyond the scope of the present work, and to our knowledge is not available in the literature. The next-to-best thing is to try to estimate the approximate average magnitude of the fields during the transition. Below, we make three attempts at such an estimate based on different lines of reasoning.

First, in ref.~\cite{ATHindmarsh}, the baryon asymmetry generated by a CP-violating operator of the type~\eqref{FFtilde} was computed as a function of the speed of transition $1/\tau_Q$, with $\tau_Q$ being the quench time, that is, the time during which the mass parameter of the Higgs field flips its sign. The generated asymmetry turns out to be largest for $m_H\tau_Q\lesssim 20$, where $m_H$ is the Higgs mass, corresponding to a characteristic momentum scale of $k_0\simeq 6\text{--}7\unit{GeV}$. Our first rough estimate of the gauge field amplitude during the spinodal transition is that of a similar order of magnitude, say $1\text{--}10\unit{GeV}$.

In the same paper~\cite{ATHindmarsh}, the $\mathrm{SU(2)}$ magnetic field was also computed as a function of time, as can be seen in figure~2 therein. Depending on the quench time and the time during the evolution at which we expect CP violation to be effective and the asymmetry to be generated, the magnitude of the field was found to be smaller than, or of the order of, the Higgs mass $m_H\simeq100\unit{GeV}$, and decreasing with the quench time. Even for the slowest quench times shown there, the magnitude of the field is still considerably larger than the above estimated scale, $\Lambda=2.6\unit{GeV}$, at $T_\text{eff}=1\unit{GeV}$. 

Thirdly, we can use the results of ref.~\cite{wmass} to obtain a more quantitative estimate of the magnitude of the gauge fields. In that work, numerical simulations were performed within the somewhat simpler $\mathrm{SU(2)}$-Higgs model, and the particle numbers of the $\mathrm{SU(2)}$ gauge field $A^a_\mu$ were computed. For the present discussion, we will assume that this gauge field is representative of the field $W^\pm_\mu$ used in this paper. Accidentally, the computation was done in the unitary gauge suitable for comparison with our results here, but only for the fastest quench, $\tau_Q=0$. The particle number can be converted to a correlator 
\begin{equation}
\langle A_i^a(t,\vek x)A_j^b(t,\vek x)\rangle=\delta^{ab}\delta_{ij}\int\frac{\dd^3\vek k}{(2\pi)^3}\left(1+\frac{|\vek k|^2}{3m_W^2}\right)\frac{n_{\vek k}^T}{\omega_{\vek k}^T},
\end{equation}
where $\omega_{\vek k}^T$ is the quasiparticle dispersion relation and $n_{\vek k}^T$ the corresponding occupation number. (The superscript ``$T$'' here refers to the transverse modes of the gauge fields.) Taking for simplicity $\omega_{\vek k}^T=\sqrt{\vek k^2+m_W^2}$ and representing the occupation number by $n_{\vek k}^T=10e^{-3.45|\vek k|/m_H}$, as appropriate for the earliest time shown in the left panel of figure~7 in ref.~\cite{wmass}, we get
\begin{equation}
\langle A_i^a(t,\vek x)A_j^b(t,\vek x)\rangle\simeq\delta^{ab}\delta_{ij}(30\unit{GeV})^2.
\end{equation}

Collecting the above three estimates for the magnitude of the gauge fields, and using the value of the scale $\Lambda$ at $T_\text{eff}=1\unit{GeV}$, we find that the ratio of the sizes of the typical type 6+2 operator and the type 4+2 operator is of the order
\begin{equation}
\left(\frac{(1\text{--}10)\unit{GeV}}{2.6\unit{GeV}}\right)^2\simeq0.15\text{--}15,\qquad 
\left(\frac{\lesssim100\unit{GeV}}{2.6\unit{GeV}}\right)^2\lesssim1500,\qquad
\left(\frac{30\unit{GeV}}{2.6\unit{GeV}}\right)^2\simeq130.
\end{equation}
The third estimate is presumably the most reliable one, but it applies only to an instantaneous quench. We conclude that for all our estimates based on existing simulations of the spinodal transition, the gradient expansion is sure to fail. This means that for our computation to be reliable in this context, the transition must be very slow indeed. A slow transition could in turn be in conflict with the need for sufficient departure from equilibrium. This remains to be seen.

\subsection{Lepton sector}
\label{sec:leptons}

\begin{figure}[t]
\begin{center}
\includegraphics[width=0.7\textwidth]{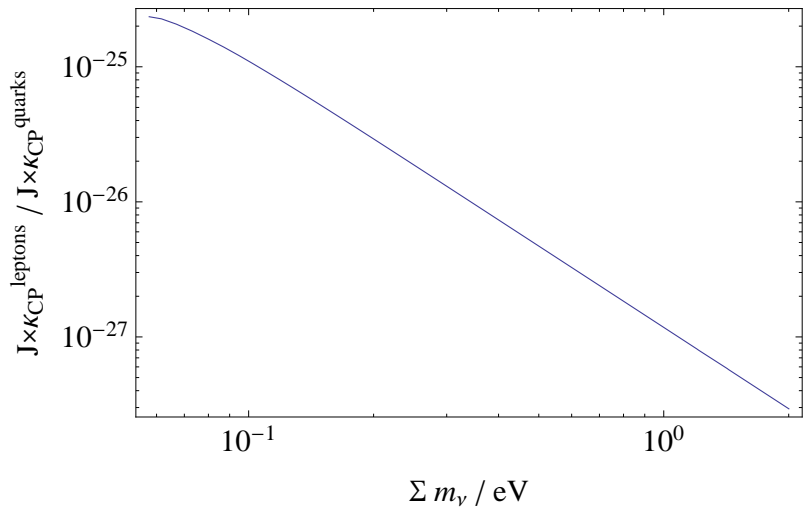}
\caption{The ratio of the zero-temperature coefficient $J\kappa_\text{CP}^{\text{4+2}}$ in the lepton sector relative to the quarks, for the whole allowed range of neutrino mass sum. The PMNS CP-violating phase is assumed maximal.}
\label{fig1:leptons2}
\end{center}
\end{figure}

As long as neutrinos can be treated as Dirac fermions, i.e., they possess no Majorana-type mass terms, a computation analogous to the above can be performed also in the lepton sector. Moreover, as there is no direct coupling between leptons and the strongly interacting sector, the calculation should be reliable all the way to zero temperature. Since the leading-order result is available from eq.~\eqref{action1}, we merely need to plug in the appropriately modified values of the input parameters to evaluate the contribution of the lepton sector to CP violation. 

In the following, we take $m_e=511\unit{keV}$, $m_\mu=106\unit{MeV}$ and $m_\tau=1.78\unit{GeV}$ for the charged lepton masses~\cite{PDG}. The absolute values of the neutrino masses are unknown, but using the fact that two differences of neutrino masses squared have been determined in neutrino oscillation experiments, $\Delta m^2_{21}=7.50\times 10^{-5}\unit{eV}^2$ and $|\Delta m^2_{32}|=2.32\times 10^{-3}\unit{eV}^2$, we can perform the calculation treating only the sum of neutrino masses as an unknown parameter. Finally, for the mixing angles in the lepton sector we take $\theta_{12}=0.59$, $\theta_{23}=0.71$ and $\theta_{13}=0.16$~\cite{PDG}. The CP-violating complex phase in the Pontecorvo-Maki-Nakagawa-Sakata (PMNS) mixing matrix is completely unknown; our approach is to fix it in order to maximize the lepton Jarlskog invariant, $\delta=\pi/2$. 

In figure~\ref{fig1:leptons2} we show the zero-temperature coefficient $J\kappa_\text{CP}^{\text{4+2}}$ for the lepton sector, normalized to the same quantity in the quark sector, for a range of the neutrino mass sum up to $2\unit{eV}$,  roughly corresponding to the upper limit given by cosmological observations. The CP-violating coupling exhibits a clear power-law scaling with the neutrino mass sum, which is easy to understand. As long as the neutrino mass sum is much larger than the mass differences, neutrinos are almost degenerate with a common mass, $m_\nu$. Very much like in the quark sector, the momentum integral in eq.~\eqref{kappaCP} can be evaluated approximately by factorizing out the masses of the heaviest contributing fermions, this time the charged leptons, leading to the simple expression
\begin{equation}
\kappa_\text{CP}^\text{4+2}\approx\frac{\Delta m_{21}^2(\Delta m_{32}^2)^2}{80\pi^2G_\text{F}m_\nu^2m_e^4m_\mu^2}.
\end{equation}

The temperature dependence of the coefficient can be computed exactly the same way as for quarks. We find that the temperature suppression is even stronger for leptons and happens at much lower temperatures due to the fact that the lepton masses are so small. From the extremely small values of the lepton $J\kappa_\text{CP}^{\text{4+2}}$, we conclude that their contribution to effective CP violation is negligible compared to the quark sector. For technical reasons related to the setup of our computational approach, this conclusion does not immediately apply if neutrinos have additional Majorana masses. However, we consider it likely that only extremely large Majorana masses would be able to compensate for the temperature suppression witnessed above. 

\subsection{Checking the results}
\label{sec:checks}

Although a thorough cross-checking of the results prior to publication is a standard part of scientific work, we find it worthwhile to discuss it in some detail here. This is due to two reasons. First, we used a computer code that performs most of the calculations without human intervention, and while this eliminates the possibility of a trivial manipulation error, it may leave a feeling of a ``black box'', the output of which must be analyzed particularly carefully. Second and more importantly, the description of our checking routines will provide the reader with additional insight into the structure of the result and its underlying symmetries.

The leading order CP-violating effective action at zero temperature was originally computed in ref.~\cite{salcedo6}, where the operators not containing derivatives of the Higgs field were derived manually. Apart from verifying the consistency with this calculation, we performed two nontrivial cross-checks of our results. The first is based on $ud$-parity~\cite{salcedo6}, an artificial discrete symmetry, under which the labels $u,d$ of the fields as well as quark mass matrices are exchanged together with the replacement $W^\pm_\mu\to W^\mp_\mu$ and $Z_\mu\to-Z_\mu$. Such an operation can be compensated by a similarity transformation on the Dirac operator~\eqref{eq:SMmat}, so that it must leave the effective action invariant. On the formal level of eq.~\eqref{master_covariant}, it corresponds to a cyclic permutation of the terms in the trace. Therefore, verifying the $ud$-parity invariance of the result is a nontrivial check that the method of (covariant) symbols has been correctly implemented.\footnote{Note that while cyclicity is a property of the full functional trace, it can be lost in the intermediate stages of the computation corresponding to eqs.~\eqref{symbols} or~\eqref{csymbols}, as the trace is performed only over the subspace of internal degrees of freedom in these expressions.}

Also the second, more nontrivial check can be formally understood as testing the implementation of the method of (covariant) symbols against a trace property that must be automatically satisfied. In contrast to cyclicity, this is the invariance of the trace under similarity transformations. As is obvious from eq.~\eqref{effactiongeneral}, both parity components of the effective action must be invariant under the transformation $K\to UKU^{-1}$ as long as $U$ commutes with $\gamma_5$. A particularly interesting choice is $U=\phi/v$, which commutes with everything inside $K$ except for the covariant derivatives, which transform as $UD_\mu^{u,d}U^{-1}=D_\mu^{u,d}+\phi[D_\mu^{u,d},\phi^{-1}]=D_\mu^{u,d}-\varphi_\mu$. This shift of the covariant derivatives dramatically reduces the dependence of the intermediate result on the Higgs field, as it disappears from the offdiagonal components of the $K$-operator~\eqref{Kop} and thus no longer explicitly appears in eq.~\eqref{master_covariant}. While this similarity transformation only affects the dependence of the intermediate expressions on the Higgs field, it does constitute of a nontrivial test of the most complicated part of the calculation. As already remarked above, once the Higgs derivatives $\varphi_\mu$ are discarded, much of the computation can be carried out manually. Why this is so can be seen from the sheer number of $\varphi$-dependent terms appearing in the expansion coefficients~\eqref{Ucoeff}.

\section{Applications}
\label{sec:appl}

\subsection{Implications for baryogenesis}
\label{sec:baryo}

The ultimate goal of our calculation was to pin down to what extent standard model CP violation can contribute to baryogenesis. Our result is particularly useful in this regard, since it not only provides a dimensionless measure of the strength of CP breaking (the coefficients $c_i$), but also an effective, purely bosonic, action in terms of these coefficients and the corresponding operators. 

The computation described above was performed assuming thermal equilibrium, and is as such well defined up to the issues related to the gradient expansion. In a general real time out-of-equilibrium context, other operators can (and will) be generated, but as the system thermalizes, the result should return to the (Wick rotated version of the) equilibrium limit.

As we have seen, at temperatures above roughly $10\unit{GeV}$, effective CP violation is suppressed by $10^{-14}$ or more compared to the zero-temperature value, which leads us to conclude that in a high-temperature electroweak first-order phase transition, standard model CP violation is of negligible magnitude. Although the details of the calculation may change in the specific context of an expanding bubble wall interacting with a hot plasma, these conditions are unlikely to alter our conclusion, in particular at $T\simeq 100\unit{GeV}$. At low temperatures, $T\simeq 1\unit{GeV}$, the picture is less clear, as we will describe below. 

A temperature as low as $1\unit{GeV}$ for electroweak baryogenesis may be achieved if the universe supercools way below the electroweak scale prior to electroweak symmetry breaking. This scenario of cold electroweak baryogenesis has been advocated for some time~\cite{cewbag1,cewbag2,cewbag3,cewbag4}, and can be realized if the Higgs field is coupled to another scalar, which may~\cite{cewbag3,lowT1} or may not~\cite{lowT2} be the inflaton, or if the first order electroweak transition takes place at a low temperature because the Higgs potential has a non-standard form~\cite{lowT3,lowT4}.

\subsection{Matching to past simulations}
\label{sec:simulations}

As mentioned above, the cold electroweak baryogenesis scenario has been studied in fully nonperturbative lattice simulations of bosonized effective theories, including $\mathrm{SU}(2)$ gauge fields, a Higgs scalar field and effective CP- and P-violating terms. The simplest such term,
\begin{equation}
\delta\Gamma_\text{CP-odd}\equiv\int_{\fv x}\frac{3\delta_\text{CP}}{16\pi^2 m_W^2}\phi^\dagger\phi\tr F^{\mu\nu}\tilde{F}_{\mu\nu},
\label{FFtilde}
\end{equation}
where $F_{\mu\nu}$ is the field strength tensor of the $\mathrm{SU}(2)$ group and $\phi$ stands for the full Higgs doublet here, was considered in refs.~\cite{cewbag4,ATSmitCP,ATHindmarsh}, where it was found that the observed baryon asymmetry is reproduced for $\delta_\text{CP}\simeq J$~\cite{cewbag4}. Such an operator, however, does not appear in the standard model~\cite{smit4}, at least in a gradient expansion of the fermion determinant.

In ref.~\cite{tranberg1} simulations were performed with an operator of type 4+2, albeit of the type $\mathcal{O}_{1}^-$, which --- as we have demonstrated --- is also absent from the standard model. Being based on a computation at zero temperature~\cite{schmidt6}, the simulations did not take into account the $\phi$-dependence of the (analogue of the) $c_i$'s and hence diverged at $\phi=0$. To bypass this issue, a cutoff procedure was introduced~\cite{tranberg2}, resulting in the asymmetry
\begin{equation}
\frac{n_\text{B}}{n_\gamma}\simeq2.4\times 10^{-6}\times\frac{\tilde{\kappa}_\text{CP}^\text{4+2}}{9.87}\exp\left(-\frac{\Lambda-50\unit{GeV}}{20\unit{GeV}}\right),
\end{equation}
which upon correction for different normalizations of $\kappa_\text{CP}^\text{4+2}$ here and in refs.~\cite{tranberg1,tranberg2} roughly translates to (assuming the appropriate cutoff is around $\Lambda=50\unit{GeV}$, as argued in ref.~\cite{tranberg2})
\begin{equation}
\frac{n_\text{B}}{n_\gamma}\simeq 1.5\times 10^{-5}\times  c_{i},
\end{equation}
where $c_i$ represents a generic coefficient of a 4+2 term. As the simulations using the full action~\eqref{action1} derived here are yet to be done, we make the very naive assumption that the asymmetry generated from the $\mathcal O^-_1$-type term is comparable to the asymmetry produced by the P-even terms of eq.~\eqref{action1}. The observed baryon asymmetry, eq.~\eqref{obsasym}, can then be reproduced provided that $c_{i}\simeq 4\times 10^{-5}$, which roughly corresponds to $T_\text{eff}\simeq1\unit{GeV}$. This means that successful baryogenesis is possible if the effective temperature at the time when baryon number violation is active is around or below $1\unit{GeV}$, with an added contribution from each operator in the action~\eqref{action1}.

In a cold spinodal electroweak transition, the fermion fields are far from equilibrium, and we should thus strictly speaking redo the computation of the $c_i$'s in the correct out-of-equilibrium state. As it is not clear how this can be achieved in practice, one might instead be interested in estimating an \emph{effective} temperature, to gauge the strength of CP violation during the transition. This effective temperature corresponding to a given out-of-equilibrium state would then be whatever gives the same value for the $c_i$ coefficients.\footnote{Note that for very out-of-equilibrium states, this may be hard to achieve as the different $c_i$ include linear combinations of different moments of the momentum distribution. Reproducing all of the $c_i$ simultaneously by a thermal distribution is nontrivial. But considering a single $c_i$ and a single operator, this can be envisaged.}

In line with the discussion of convergence of the gradient expansion in section~\ref{sec:order8}, one possible estimate of the momentum scale involved, and thus of the effective temperature, is obtained by translating the quench time $\tau_Q$ into a frequency, $k_0\simeq \tau_Q^{-1}$. As mentioned there, we know from ref.~\cite{ATHindmarsh} that in order to get a sizable asymmetry, the quench should not be slower than $\tau_Qm_H\simeq20$, thus leading to $k_0\simeq6\text{--}7\unit{GeV}$. As we have seen that the effective CP violation is extremely sensitive to temperature, a more precise estimate is clearly necessary. Ultimately, the use of the bosonic effective action based on the gradient expansion should be replaced by first-principle simulations of the electroweak transition with dynamical fermions. In the following, we will discuss the future prospects for this.

\subsection{Future bosonic simulations}
\label{sec:bosonsim}

Baryon number is C- and CP-odd, and P-even. Chern-Simons number is P- and CP-odd, but C-even. The chiral anomaly, which is a direct result of C- and P-breaking in the fermion-gauge interactions, relates the two through
\begin{equation}
B(t)-B(0)=3[N_\text{CS}(t)-N_\text{CS}(0)],
\end{equation}
and in order to generate a baryon asymmetry, they must therefore both become non-zero. This is only possible if all the discrete symmetries C, P and CP are broken.

In a purely bosonic simulation, baryon number is not explicit, but assumed to follow the evolution of the Chern-Simons number. As a consequence, it is necessary to include the effects of both P and CP violation to generate non-zero asymmetry. Had the outcome of our computation of the effective action been CP violation in the P-breaking sector, the computed operators would be sufficient, as demonstrated in~\cite{tranberg1, tranberg2}. But it has now been convincingly demonstrated~\cite{salcedo6,PRL} that only CP violation in the C-breaking sector appears at leading order, and therefore bosonic simulations including only this would still give zero net Chern-Simons number.

The way out is to combine the P-even and C-odd operators from eq.~\eqref{action1} with the CP-conserving, C-odd and P-odd terms which originate from the left-handed gauge couplings to fermions. These should appear already at the fourth order of the gradient expansion and are naturally not proportional to the Jarlskog invariant. We have not computed these terms here, but the method of symbols is ideally suited for doing so, although complications may arise at this order through the Wess-Zumino-Witten term.

As an aside, a similar complication arises in the two-Higgs doublet model. There, CP violation is provided at tree level through the complex couplings in the two-scalar potential, which break CP by breaking charge conjugation and preserving parity. Since CP violation is now located in the scalar potential rather than the fermion Yukawa couplings, one might expect to be able to generate an asymmetry using the bosonic fields only. However, parity is conserved, so the Chern-Simons number remains zero. One thus again has to include the C-odd and P-odd operators that allow for the generation of the Chern-Simons number. 

Simulations of this setup were recently carried out in ref.~\cite{BinWu}, showing that an asymmetry is indeed created, but only when CP violation is combined with the C/P-violating operator(s). However, rather than explicitly computing the C- and P-breaking terms coming from the gauge couplings to fermions, the authors simply assume an effective operator of the type
\begin{equation}
\delta\Gamma_{\text{C-odd/P-odd}}=\imag\kappa_\text{C/P}\bigl(\phi_1\phi_2^\dagger-\phi_2\phi_1^\dagger\bigr)\tr F_{\mu\nu}\tilde{F}^{\mu\nu},
\label{eq:2HDM}
\end{equation}
related to the chiral anomaly~\cite{Turok}. Here $\phi_{1,2}$ are the two Higgs doublets. 

\subsection{Future simulations including fermions: maximizing CP violation}
\label{sec:maximizing}

\begin{figure}[t]
\begin{center}
\includegraphics[width=0.7\textwidth]{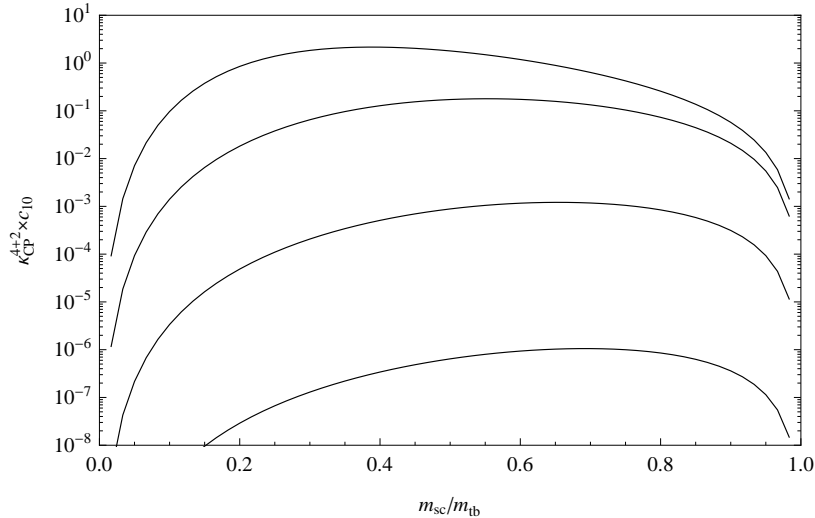}
\caption{The coefficient $\kappa_\text{CP}^\text{4+2}c_{10}$ at $T_\text{eff}=1\unit{GeV}$ for $m_{ud}=0$ as a function of $m_{sc}/m_{tb}$ for $m_{tb}/T_\text{eff}=1,2,4,8$. The successive values correspond to the plotted lines from the bottom to the top.}
\label{fig:cpopt1}
\end{center}
\end{figure}

Ultimately, one would like to perform simulations including dynamical fermions in real time, along the way described in refs.~\cite{PF1,PF2}. This bypasses the use of the gradient expansion, and the fermions will automatically be in the correct out-of-equilibrium state. On the other hand, the numerical effort is huge, since the method requires stochastic averaging over fermion field ensembles for each member of a bosonic field ensemble. 

The observed baryon asymmetry is $n_\text{B}/n_\gamma\simeq 10^{-9}$, and such a signal is very hard to see in numerical Monte-Carlo simulations. In practice, what is done is to simulate the dynamics with a value for $\delta_\text{CP}$ (of whatever bosonic operator) much larger than the physical one, find the regime where the baryon asymmetry is linearly dependent on $\delta_\text{CP}$, and interpolate the result to the physical value. For purely bosonic simulations, one can in principle choose $\delta_\text{CP}$ to have any value. When including fermions, the CP violation is however encoded in the actual CKM matrix, the structure of which does not allow arbitrarily large CP violation. What we can do is try to maximize the baryon asymmetry by modifying mixing angles, the CP-violating phase and the fermion masses. Although we do not a priori know how the asymmetry depends on these parameters, we can get a good estimate of the optimal point in parameter space by maximizing $\kappa_\text{CP}^\text{4+2}$ and the $c_i$. 

First of all, CP violation is generically proportional to the Jarlskog invariant, which takes its maximum for the (unphysical, but in principle allowed) mixing angles $\sin\theta_{12}=1/\sqrt{2}$, $\sin\theta_{13}=1/\sqrt{3}$, $\sin\theta_{23}=1/\sqrt{2}$ and the complex CP-violating phase $\delta=\pi/2$. The resulting value $J=1/(6\sqrt3)\approx 0.096$ is larger than the physical value by the factor of $3.3\times 10^{3}$. 

Second, the CP-violating effective couplings depend in a particular, and rather nontrivial, way on the quark masses. The overall factor $\Delta$ is simply maximized for large mass separations, but the complete expression is more complicated. We start by simplifying the problem by taking the $u$-type and $d$-type quark masses to be equal, $m_u=m_d\equiv m_{ud}$, $m_c=m_s\equiv m_{sc}$ and $m_t=m_b\equiv m_{tb}$. In addition, we will take the limit $m_{ud}\rightarrow 0$ since the actual value of this mass plays very little role; $\kappa_\text{CP}^\text{4+2}$ is regular and does not change significantly in this limit. Consequently, the coefficients $\kappa_\text{CP}^\text{4+2}c_i$ depend on three independent scales: $m_{sc}$, $m_{tb}$ and the effective temperature $T_\text{eff}$.

By simple dimensional analysis, $\kappa_\text{CP}^\text{4+2}c_i$ can be written as $1/T_\text{eff}^2$ times a function of two dimensionless ratios, which can be chosen conveniently as $m_{sc}/m_{tb}$ and $m_{tb}/T_\text{eff}$. While the former characterizes the fermion mass hierarchy, the latter compares the overall mass scale to the temperature. Figure~\ref{fig:cpopt1} displays the largest of the coefficients, $\kappa_\text{CP}^\text{4+2}c_{10}$, as a function of the mass hierarchy $m_{sc}/m_{tb}$ for several values of $m_{tb}/T_\text{eff}$. While the plot applies directly to the case of $T_\text{eff}=1\unit{GeV}$, the values of the coupling at other temperatures can be obtained by mere rescaling. 

Varying the three available parameters, $T_\text{eff}$, $m_{sc}$ and $m_{tb}$, creates a lot of space for enhancement of the signal for CP violation. For instance, by choosing $T_\text{eff}=1\unit{GeV}$ and $m_{tb}=4\unit{GeV}$, figure~\ref{fig:cpopt1} tells us that we should tune $m_{sc}$ so that $m_{sc}/m_{tb}\approx0.6$. The coefficient $\kappa_\text{CP}^\text{4+2}c_{10}$ then reaches the value of roughly $10^{-1}$, which is to be compared to its value at the physical point, $3\times10^{-3}$. While this is merely a concrete example of the dependence of the effective coupling on the parameters as displayed in figure~\ref{fig:cpopt1}, additional constraints due to the setup of the numerical simulations need to be taken into account. The fermion lattice simulations in refs.~\cite{PF1,PF2} acquire much larger statistical uncertainties for large fermion masses, so we would like to keep $m_{sc}<m_{tb}\lesssim10\unit{GeV}$. On the other hand, we also need to resolve the physical gauge and Higgs boson masses on the lattice, so the fermions should not be too light, either. 

By combining the above concrete choice of fermion masses with the optimization of the Jarlskog invariant, we are able to increase the expected asymmetry by a factor of at least $10^5$. Conversely, if one performed simulations including fermions with these optimized parameters, then finding a net asymmetry of the order of $10^{-5}$ should correspond to the observed asymmetry at the physical point. Note that this is precisely the order of magnitude of the asymmetry seen in bosonic simulations (see for instance ref.~\cite{ATSmitCP}). In other words, although the scope to tune CP violation is limited to only shifting masses and mixings rather than freely tuning $\delta_\text{CP}$, the numerical signal can still be made large enough to be seen in a real-time Monte Carlo lattice simulation with numerically manageable statistics.

\section{Conclusions}
\label{sec:conclusion}

Integrating out fermions from the standard model gives rise to effective CP-violating operators for the gauge and Higgs fields. In the present paper, we have shown how one can obtain such operators systematically within a covariant gradient expansion, at both zero and nonzero temperature. We have computed the temperature dependence of the effective couplings at the leading nontrivial, sixth order in the gradient expansion, and found that to this order, there are no operators that simultaneously violate both the CP and P symmetries. Although detailed numerical simulations using the resulting bosonic effective action are yet to be performed, a simple estimate allowed us to conclude that standard model CP violation may be the source of the cosmological baryon asymmetry provided that the electroweak phase transition happens at low enough temperature, of the order of $1\unit{GeV}$. This implies that while ``hot'' electroweak baryogenesis with the standard model as the only source of CP violation is firmly ruled out (as has been known for many years), other cosmological scenarios such as cold electroweak baryogenesis might provide a viable solution to the baryon asymmetry problem.

With regards to numerical simulations of baryogenesis, we made several observations that will partially direct our future efforts. First, the contribution of the lepton sector to CP violation can be safely neglected as long as neutrinos are of the Dirac type. The discussion of Majorana neutrinos goes beyond the scope of the present paper and will require a proper extension of the formalism used here, but we suspect that our qualitative findings will not be altered. Second, numerical bosonic simulations determine the generation of the gauge field Chern-Simons number, which is related to baryon number via the chiral anomaly. As the Chern-Simons number is odd under parity, and all of the operators we have found in the leading order CP-violating effective action preserve parity, the knowledge of simultaneously P-odd and C-odd operators is required. These arise from the purely left-handed gauge interactions of fermions and are expected to appear already at the fourth order of the gradient expansion. 

As a word of caution, and based on existing numerical simulations of spinodal electroweak symmetry breaking, we noted that the gradient expansion fails to converge unless the speed of the transition is very small. Although this neither questions the correctness of our present results nor invalidates the ``cold'' electroweak baryogenesis scenario, it raises questions regarding the applicability of the gradient expansion in this context.

As a final technical point, we have estimated the optimal values of fermion masses and mixing angles to give the maximal baryon asymmetry. This is important for future simulations of baryogenesis with dynamical fermions and the actual CKM matrix structure, since we then do not have the freedom to make the asymmetry arbitrarily large (and numerically detectable) by choosing a large $\delta_\text{CP}$ freely. 

\acknowledgments We thank L.~L.~Salcedo for many useful comments in the initial stages of this work, and NordForsk for support. T.B., O.T.~and A.V.~were funded by the Sofja Kovalevskaja program of the Alexander von Humboldt Foundation, and A.T.~by the Carlsberg Foundation.

\appendix

\section{Notation and conventions}
\label{sec:notationSM}

Throughout the paper, we use the natural units, in which the Planck and Boltzmann constants as well as the speed of light are equal to one. We also work in Euclidean space; Euclidean spacetime vectors are denoted by lowercase italic letters and their spatial parts in boldface, for instance $\fv v=(v_0,\vek v)$. Spacetime indices are represented by lowercase Greek letters; repeated indices are summed over unless explicitly stated otherwise. In the Matsubara formalism, the temporal component of momentum is discrete. For fermions, it takes the values $\omega_n\equiv (2n+1)\pi T$, where $T$ is the temperature.

In order to simplify the notation, we adopt the following conventions for integration symbols. The integration over spacetime coordinate $x$ is denoted as $\int_{\fv x}\equiv\int\dd^d\fv x$ at zero temperature, where $d$ is the dimension of the spacetime. At nonzero temperature, it has to be modified to $\int_{\fv x}\equiv\int_0^{1/T}\dd x_0\int\dd^{d-1}\vek x$. Momentum integration is defined as $\int_{\fv p}\equiv\int\dd^d\fv p/(2\pi)^d$ at zero temperature, and $\int_{\fv p}\equiv T\sum_{p_0}\int\dd^{d-1}\vek p/(2\pi)^{d-1}$ at nonzero temperature. Here, the sum is over the Matsubara frequencies, $p_0=\omega_n$, as defined above.

\subsection{Standard model}

In order to simplify the lengthy expressions we have to deal with, it is crucial to introduce a suitable notation. We follow ref.~\cite{salcedo6} and absorb the gauge couplings into the redefinition of the gauge fields. Specifically, we trade the physical fields $\tilde W^\pm_\mu$ (charged weak intermediate bosons), $\tilde Z_\mu$ (neutral weak intermediate boson), $\tilde B_\mu$ (hypercharge gauge field), and $\tilde G_{a\mu}$ (gluon color-octet field) for
\begin{equation}
W^\pm_\mu\equiv\frac g{\sqrt2}\tilde W^\pm_\mu,\qquad
Z_\mu\equiv\frac g{2\cos\theta_{\text W}}\tilde Z_\mu,\qquad
B_\mu\equiv g'\tilde B_\mu,\qquad
G_\mu\equiv\frac{g_{\text s}}2\lambda_a\tilde G_{a\mu}.
\end{equation}
Here, as usual, $g,g',g_{\text s}$ denote, respectively, the weak isospin, hypercharge, and strong coupling constants, while $\theta_{\text W}$ is the Weinberg angle and $\lambda_a$ are the Gell-Mann matrices. With this notation, the chiral components of the (Euclidean) quark Dirac operator~\eqref{eq:origD} in the unitary gauge become
\begin{equation}
\begin{split}
D_{\text L,\mu}&=\begin{pmatrix}
D_{u,\mu}+Z_\mu+G_\mu & W^+_\mu\\
W^-_\mu & D_{d,\mu}-Z_\mu+G_\mu 
\end{pmatrix},\qquad
m_{\text{LR}}=\begin{pmatrix}
\frac\phi vM_u & 0 \\
0 & \frac\phi vM_d
\end{pmatrix},\\
D_{\text R,\mu}&=\begin{pmatrix}
D_{u,\mu}+G_\mu & 0\\
0 & D_{d,\mu}+G_\mu 
\end{pmatrix},\qquad
m_{\text{RL}}=\begin{pmatrix}
\frac\phi v M^\dagger_u & 0 \\
0 & \frac\phi vM^\dagger_d
\end{pmatrix},
\end{split}
\label{SMdirac}
\end{equation}
where the matrix structure corresponds to isospin space. The hypercharge covariant derivatives used here are defined as $D_{u,\mu}\equiv\partial_\mu+(2/3)B_\mu$ and $D_{d,\mu}=\partial_\mu-(1/3)B_\mu$. In the two-dimensional space of eletrically neutral electroweak gauge fields, it is common to work with the basis of the $Z_\mu$ field and the photon field, or with that of the neutral isospin gauge field and the hypercharge field $B_\mu$. We choose here a ``mixed'' basis, composed of $Z_\mu$ and $B_\mu$. This is the reason why $Z_\mu$ does not appear in the right-handed component of the Dirac operator, and why the couplings of the hypercharge field $B_\mu$ in the covariant derivatives equal the electric charge of the fermions rather than their actual hypercharge.

For the sake of completeness, we also write down the analogous expressions in the lepton sector,
\begin{equation}
\begin{split}
D_{\text L,\mu}&=\begin{pmatrix}
\partial_\mu+Z_\mu & W^+_\mu\\
W^-_\mu & D_{e,\mu}-Z_\mu 
\end{pmatrix},\qquad
m_{\text{LR}}=\begin{pmatrix}
\frac\phi vM_\nu & 0 \\
0 & \frac\phi vM_e
\end{pmatrix}\\
D_{\text R,\mu}&=\begin{pmatrix}
\partial_\mu & 0\\
0 & D_{e,\mu} 
\end{pmatrix},\qquad
m_{\text{RL}}=\begin{pmatrix}
\frac\phi v M^\dagger_\nu & 0 \\
0 & \frac\phi vM^\dagger_e
\end{pmatrix},
\end{split}
\label{SMdiraclepton}
\end{equation}
where we denoted $D_{e,\mu}\equiv\partial_\mu-B_\mu$ and assumed that the neutrinos have only a Dirac-type mass matrix, $M_\nu$.

\section{Method of symbols}
\label{sec:symbols}

The method of symbols is a systematic device to calculate traces of differential operators. While its basic form has been known for a long time~\cite{Nepomechie} and its covariant version was introduced more than a decade ago~\cite{Pletnev}, the generalization to finite temperature has only been worked out recently~\cite{salcedoT}. Here we provide an overview of the method, following closely the notation of ref.~\cite{salcedo6}.

Our task is to calculate the trace of a differential operator $f(D,M)$ where $D_\mu$ is a covariant derivative and $M(\fv x)$ a (local) background field. Both of them can in principle have some additional matrix structure (internal degrees of freedom). In order to calculate the trace we need to choose a basis. It is convenient to choose a direct product basis, composed of independent bases in the space of test functions of the coordinate $\fv x$, denoted as $\chi_n(\fv x)$, and in the internal space. Provided the basis $\chi_n(\fv x)$ is orthonormal, the full trace becomes
\begin{equation}
\Tr f(D,M)=\sum_n\int_{\fv x}\tr\bigl[\chi_n^*(\fv x)f(D,M)\chi_n(\fv x)\bigr],
\end{equation}
where ``$\tr$'' refers to a trace over the internal space only. Choosing in particular the basis of plane waves, $\chi_{\fv p}(\fv x)\equiv e^{\imag\fv p\cdot\fv x}$, we replace the sum over $n$ with an integral over $\fv p$ and use the operator identity $e^{-\imag\fv p\cdot\fv x}De^{\imag\fv p\cdot\fv x}=D+\imag\fv p$ to obtain
\begin{equation}
\Tr f(D,M)=\int_{\fv x,\fv p}\tr\bigl[f(D+\imag\fv p,M)\openone\bigr],
\label{symbols}
\end{equation}
where $\openone$ is a unit matrix in the internal space. This formula constitutes the main result of the method of symbols.

While the above expression is fully general, its obvious disadvantage is the lack of manifest gauge covariance due to the appearance of the ``free'' covariant derivative operators~\cite{salcedo6}. In order to remedy this, let us first introduce an arbitrary vector operator $\Xi_\mu$ acting in $\fv x$-space and rewrite eq.~\eqref{symbols} as
\begin{equation}
\Tr f(D,M)=\int_{\fv x,\fv p}\tr\bigl[e^{-\imag\Xi\cdot\nabla}e^{\imag\Xi\cdot\nabla}e^{-\imag\fv p\cdot\fv x}f(D,M)e^{\imag\fv p\cdot\fv x}e^{-\imag\Xi\cdot\nabla}\openone\bigr],
\end{equation}
where we used the shorthand notation $\nabla_\mu\equiv\partial/\partial\fv p_\mu$. Note that adding the exponential on the far right does not change anything since it acts on the unit matrix, while the two extra exponentials before $f(D,M)$ combine to a unity. Therefore, this is an identity that holds regardless of the temperature. 

Next, we define a barred symbol for an operator $v$ through
\begin{eqnarray}
\label{eq:similarity}
\bar v\equiv e^{\imag\Xi\cdot\nabla}e^{-\imag\fv p\cdot\fv x}\,v\,e^{\imag\fv p\cdot\fv x}e^{-\imag\Xi\cdot\nabla}.
\label{vbar}
\end{eqnarray}
As the last step, we expand the first exponential $e^{-\imag\Xi\cdot\nabla}$ in powers of the derivative and note that its spatial part produces zero, as it gives an integral of a total derivative in momentum space. (This observation relies on the fact that $\Xi_\mu$ acts in coordinate space only and thus commutes with $\nabla_\mu$.) The final result for the trace of the differential operator in the method of covariant symbols therefore reads
\begin{equation}
\Tr f(D,M)=\sum_{n=0}^\infty\frac{(-\imag)^n}{n!}\int_{\fv x,\fv p}\tr\bigl[(\Xi_0\nabla_0)^nf(\bar D,\bar M)\openone\bigr].
\label{csymbols}
\end{equation}
Note that at zero temperature only the $n=0$ term of the series survives, and in addition that we have complete freedom in the choice of the operator $\Xi_\mu$. It is, however, advantageous to simply set $\Xi_\mu=D_\mu$, since the ``free'' derivative in $D_\mu+\imag\fv p_\mu$ is then cancelled and $\bar D_\mu$ becomes a multiplicative operator, containing covariant derivatives only inside commutators. Concretely, the expansions of the covariant derivative and the scalar field read
\begin{equation}
\begin{split}
\bar D_\mu=e^{\imag D\cdot\nabla}(D_\mu+\imag\fv p_\mu)e^{-\imag D\cdot\nabla}
&=\imag\fv p_\mu+\sum_{n=1}^\infty\frac{\imag^n}{(n+1)(n-1)!}F_{\alpha_1\dotsb\alpha_n\mu}\nabla_{\alpha_1}\dotsb\nabla_{\alpha_n}\\
&=\imag\fv p_\mu+\frac\imag2[D_\alpha,D_\mu]\nabla_\alpha+\dotsb,\\
\bar M=e^{\imag D\cdot\nabla}Me^{-\imag D\cdot\nabla}
&=M+\sum_{n=1}^\infty\frac{\imag^n}{n!}M_{\alpha_1\dotsb\alpha_n}\nabla_{\alpha_1}\dotsb\nabla_{\alpha_n}\\
&=M+\imag[D_\alpha,M]\nabla_\alpha-\frac12[D_\alpha,[D_\beta,M]]\nabla_\alpha\nabla_\beta+\dotsb,\\
\end{split}
\label{cov_symbols}
\end{equation}
where $M_{\alpha_1\dotsb\alpha_n}\equiv[D_{\alpha_1},[\dotsb[D_{\alpha_n},M]\dotsb]]$ and $F_{\alpha_1\dotsb\alpha_n\mu}\equiv[D_{\alpha_1},[\dotsb[D_{\alpha_n},D_\mu]\dotsb]]$. This concludes the derivation of eq.~(5) presented in ref.~\cite{PRL}, and agrees with eq.~(2.40) derived independently in ref.~\cite{salcedoT}.

\subsection{Application to the standard model}
\label{sec:symbolsSM}

In order to be able to apply the method of covariant symbols to the standard model, we have to respect the isospin structure of its Dirac operator. We therefore choose
\begin{equation}
\Xi_\mu=\begin{pmatrix}
D^u_\mu & 0\\
0 & D^d_\mu
\end{pmatrix}
\end{equation}
in isospin space. Should the terms including the gluon field be desired as well, we would simply replace $D^{u,d}_\mu$ with $D^{u,d}_\mu+G_\mu$ in the above. As the gluon field commutes with all other fields but the covariant derivative itself, this shift would only affect the field strength tensors $F^{u,d}_{\alpha_1\dotsb\alpha_n\mu}$ defined below, giving them a separate contribution in color space. The barred electroweak fields, defined by eq.~\eqref{vbar}, now take the form
\begin{align}
\notag
\bar D_\mu^{u,d}&=e^{\imag D^{u,d}\cdot\nabla}(D_\mu^{u,d}+\imag\fv p_\mu)e^{-\imag D^{u,d}\cdot\nabla}=\imag\fv p_\mu+\sum_{n=1}^\infty\frac{\imag^n}{(n+1)(n-1)!}F^{u,d}_{\alpha_1\dotsb\alpha_n\mu}\nabla_{\alpha_1}\dotsb\nabla_{ \alpha_n},\\
\notag\bar W_\mu^\pm&=e^{\imag D^{u,d}\cdot\nabla}W_\mu^\pm e^{-\imag D^{d,u}\cdot\nabla}=W_\mu^\pm+\sum_{n=1}^\infty\frac{\imag^n}{n!}W^\pm_{\alpha_1\dotsb\alpha_n\mu}\nabla_{\alpha_1}\dotsb\nabla_{\alpha_n},\\
\label{SMcovsymb}
\bar Z_\mu&=e^{\imag D^{u,d}\cdot\nabla}Z_\mu e^{-\imag D^{u,d}\cdot\nabla}=Z_\mu+\sum_{n=1}^\infty\frac{\imag^n}{n!}Z_{\alpha_1\dotsb\alpha_n\mu}\nabla_{\alpha_1}\dotsb\nabla_{\alpha_n},\\
\notag\bar\varphi_\mu&=e^{\imag D^{u,d}\cdot\nabla}\varphi_\mu e^{-\imag D^{u,d}\cdot\nabla}=\varphi_\mu+\sum_{n=1}^\infty\frac{\imag^n}{n!}\varphi_{\alpha_1\dotsb\alpha_n\mu}\nabla_{\alpha_1}\dotsb\nabla_{\alpha_n},
\label{covar_symb}
\end{align}
where $F^{u,d}_{\alpha_1\dotsb\alpha_n\mu}\equiv[D^{u,d}_{\alpha_1},[\dotsb[D^{u,d}_{\alpha_n},D^{u,d}_\mu ]\dotsb]]$, $Z_{\alpha_1\dotsb\alpha_n\mu}\equiv[\partial_{\alpha_1},[\dotsb[\partial_{\alpha_n},Z_\mu]\dotsb]]$, and analogously for $\varphi_{\alpha_1\dotsb\alpha_n\mu}$. Furthermore, we have defined here $W^+_{\alpha_1\alpha_2\dotsb\alpha_n\mu}\equiv D^u_{\alpha_1}W^+_{\alpha_2\dotsb\alpha_n\mu}-W^+_{\alpha_2\dotsb\alpha_n\mu}D^d_{\alpha_1}$ and $W^-_{\alpha_1\alpha_2\dotsb\alpha_n\mu}\equiv D^d_{\alpha_1}W^-_{\alpha_2\dotsb\alpha_n\mu}-W^-_{\alpha_2\dotsb\alpha_n\mu}D^u_{\alpha_1}$. Note that the above compact expression for $\bar W_\mu^\pm$ follows from the generalization of the Baker-Campbell-Hausdorff formula,
\begin{equation}
e^{A}Xe^{-B}=\exp\bigl(\bigl[\genfrac{}{}{0pt}{1}{A}{B},\,\cdot\,\bigr]\bigr)X,
\end{equation}
where $\bigl[\genfrac{}{}{0pt}{1}{A}{B},X\bigr]\equiv AX-XB$. Finally, we need the expression for $\bar\phi$,
\begin{equation}
\bar\phi=e^{\imag D^{u,d}\cdot\nabla}\phi\,e^{-\imag D^{u,d}\cdot\nabla}=\phi\biggl(1+\sum_{n=1}^\infty\frac{\imag^n}{n!}\Omega_{\alpha_1\dotsb\alpha_n}\nabla_{\alpha_1}\dotsb\nabla_{\alpha_n}\biggr),
\end{equation}
where the structure of the $\Omega$-tensor is determined by the partitions of the set of indices $\alpha_1,\dotsc,\alpha_n$,
\begin{equation}
\Omega_{\alpha_1\dotsb\alpha_n}=\sum_{\substack{\text{partitions }\beta_{ij}\\
\bigcup_i\{\beta_{ij}\}_j=\{\alpha_n\}_n}}\prod_i\varphi_{\beta_{i1}\beta_{i2}\dotsb},
\end{equation}
for example
\begin{align}
\notag
\Omega_{\alpha\beta}=&\varphi_{\alpha\beta}+\varphi_\alpha\varphi_\beta,\\
\notag
\Omega_{\alpha\beta\gamma}=&\varphi_{\alpha\beta\gamma}+\varphi_{\alpha\beta}
\varphi_\gamma+\varphi_{\alpha\gamma}\varphi_\beta+\varphi_{\beta\gamma}
\varphi_\alpha+\varphi_\alpha\varphi_\beta\varphi_\gamma,\\
\Omega_{\alpha\beta\gamma\delta}=&\varphi_{\alpha\beta\gamma\delta}
+\varphi_\alpha\varphi_\beta\varphi_\gamma\varphi_\delta\\
\notag
&+\varphi_{\alpha\beta\gamma}\varphi_\delta+\varphi_{\alpha\beta\delta}\varphi_\gamma
+\varphi_{\alpha\gamma\delta}\varphi_\beta+\varphi_{\beta\gamma\delta}\varphi_\alpha
+\varphi_{\alpha\beta}\varphi_{\gamma\delta}+\varphi_{\alpha\gamma}\varphi_{\beta\delta}
+\varphi_{\alpha\delta}\varphi_{\beta\gamma}\\
\notag
&+\varphi_{\alpha\beta}\varphi_\gamma\varphi_\delta+\varphi_{\alpha\gamma}
\varphi_\beta\varphi_\delta+\varphi_{\alpha\delta}\varphi_\beta\varphi_\gamma+\varphi_{\beta\gamma}
\varphi_\alpha\varphi_\delta+\varphi_{\beta\delta}\varphi_\alpha\varphi_\gamma+\varphi_{\gamma\delta
}\varphi_\alpha\varphi_\beta.
\end{align}

In order to be able to evaluate traces of the type~\eqref{master_covariant}, we still need to know the propagator in the method of covariant symbols, cf.~eq.~\eqref{eq:master_prop},
\begin{equation}
\bar{\tilde N}_{u,d}^{-1}=\frac{\bar\phi^2}{v^2}M_{u,d}M_{u,d}^\dagger-(\bar{\slashed D}_{u,d}\pm\bar{\slashed Z})(\bar{\slashed D}_{u,d}+\bar{\slashed\varphi}).
\end{equation}
This is expanded up to the desired order in the covariant gradient expansion, $\bar{\tilde N}^{-1}_{u,d}\equiv N^{-1}_{u,d}-(U_1^{u,d}+U_2^{u,d}+\dotsb)$, and the series is inverted so that the propagator reads
\begin{equation}
\bar{\tilde N}_{u,d}=N_{u,d}+N_{u,d}(U_1^{u,d}+U_2^{u,d})N_{u,d}+N_{u,d}U_1^{u,d}N_{u,d}U_1^{u,d}N_{u,d}+\text{higher-order terms}.
\label{Nbar}
\end{equation}
The expansion coefficients of the inverse propagator read, up to fourth order in the gradient expansion,
\begin{align}
\notag U_1^{u,d}=&-2\imag G_{u,d}\varphi_\alpha\nabla_\alpha\pm\imag\slashed{Z}\slashed{\fv p}+\imag\slashed{\fv p}\slashed\varphi,\\
\notag U_2^{u,d}=&G_{u,d}\bigl(\varphi_{\alpha\beta}+2\varphi_\alpha\varphi_\beta\bigr)\nabla_\alpha\nabla_\beta-\bigl(\tfrac12F^{u,d}_{\alpha\mu}\pm Z_{\alpha\mu}\bigr)\gamma_\mu\nabla_\alpha\slashed{\fv p}-\bigl(\tfrac12F_{\alpha\mu}^{u,d}+\varphi_{\alpha\mu}\bigr)\slashed{\fv p}\gamma_\mu\nabla_\alpha\pm\slashed Z\slashed\varphi,\\
\notag U_3^{u,d}=&\imag G_{u,d}\bigl(\tfrac13\varphi_{\alpha\beta\gamma}+2\varphi_{\alpha\beta}\varphi_\gamma+\tfrac43\varphi_\alpha\varphi_\beta\varphi_\gamma\bigr)\nabla_\alpha\nabla_\beta\nabla_\gamma-\\
\label{Ucoeff}
&-\imag\bigl(\tfrac13F^{u,d}_{\alpha\beta\mu}\pm\tfrac12Z_{\alpha\beta\mu}\bigr)\gamma_\mu\nabla_\alpha\nabla_\beta\slashed{\fv p}-\imag\bigl(\tfrac13F^{u,d}_{\alpha\beta\mu}+\tfrac12\varphi_{\alpha\beta\mu}\bigr)\slashed{\fv p}\gamma_\mu\nabla_\alpha\nabla_\beta+\\
\notag &+\imag\bigl(\tfrac12F_{\alpha\mu}^{u,d}\pm Z_{\alpha\mu}\bigr)\gamma_\mu\slashed\varphi\nabla_\alpha\pm\imag\bigl(\tfrac12F_{\alpha\mu}^{u,d}+\varphi_{\alpha\mu}\bigr)\slashed Z\gamma_\mu\nabla_\alpha,\\
\notag U_4^{u,d}=&-G_{u,d}\bigl(\tfrac1{12}\varphi_{\alpha\beta\gamma\delta}+\tfrac23\varphi_{\alpha\beta\gamma}\varphi_\delta+\tfrac12\varphi_{\alpha\beta}\varphi_{\gamma\delta}+2\varphi_{\alpha\beta}\varphi_\gamma\varphi_\delta+\tfrac23\varphi_\alpha\varphi_\beta\varphi_\gamma\varphi_\delta\bigr)\nabla_\alpha\nabla_\beta\nabla_\gamma\nabla_\delta+\\
\notag &+\bigl(\tfrac18F_{\alpha\beta\gamma\mu}^{u,d}\pm\tfrac16Z_{\alpha\beta\gamma\mu}\bigr)\gamma_\mu\nabla_\alpha\nabla_\beta\nabla_\gamma\slashed{\fv p}+\bigl(\tfrac18F_{\alpha\beta\gamma\mu}^{u,d}+\tfrac16\varphi_{\alpha\beta\gamma\mu}\bigr)\slashed{\fv p}\gamma_\mu\nabla_\alpha\nabla_\beta\nabla_\gamma-\\
\notag &-\bigl(\tfrac13F_{\alpha\beta\mu}^{u,d}\pm\tfrac12Z_{\alpha\beta\mu}\bigr)\gamma_\mu\slashed\varphi\nabla_\alpha\nabla_\beta\mp\bigl(\tfrac13F_{\alpha\beta\mu}^{u,d}+\tfrac12\varphi_{\alpha\beta\mu}\bigr)\slashed Z\gamma_\mu\nabla_\alpha\nabla_\beta-\\
\notag &-\bigl(\tfrac12F_{\alpha\mu}^{u,d}\pm Z_{\alpha\mu}\bigr)\bigl(\tfrac12F^{u,d}_{\beta\nu}+\varphi_{\beta\nu}\bigr)\gamma_\mu\gamma_\nu\nabla_\alpha\nabla_\beta,
\end{align}
where we have introduced the shorthand notation $G_{u,d}\equiv N_{u,d}^{-1}-\fv p^2$. This ensures that all momentum integrals resulting from the method of covariant symbols are of the type~\eqref{IintT}.

\section{Momentum integrals in the 4+2 sector at finite temperature}
\label{sec:integrals}

Upon diagonalization of the quark mass matrices, the momentum integrals needed in the evaluation of the 4+2 type operators in the CP-violating effective action take the form
\begin{equation}
\begin{split}
\hat I^{k,3-k}_{1,1,2,2}&=\imag\im\sum_{\substack{i,j,m,n}}J_{ijmn}\int_{\fv p}\frac{(\vek p^2)^k(p_0^2)^{3-k}}{(\fv p^2+m_{u,i}^2)(\fv p^2+m_{d,j}^2)(\fv p^2+m_{u,m}^2)^2(\fv p^2+m_{d,n}^2)^2},\\
\hat I^{k,4-k}_{1,1,2,3}&=\imag\im\sum_{\substack{i,j,m,n}}J_{ijmn}\int_{\fv p}\frac{(\vek p^2)^k(p_0^2)^{4-k}}{(\fv p^2+m_{u,i}^2)(\fv p^2+m_{d,j}^2)(\fv p^2+m_{u,m}^2)^2(\fv p^2+m_{d,n}^2)^3},\\
\hat I^{k,4-k}_{1,1,3,2}&=\imag\im\sum_{\substack{i,j,m,n}}J_{ijmn}\int_{\fv p}\frac{(\vek p^2)^k(p_0^2)^{4-k}}{(\fv p^2+m_{u,i}^2)(\fv p^2+m_{d,j}^2)(\fv p^2+m_{u,m}^2)^3(\fv p^2+m_{d,n}^2)^2},
\end{split}
\label{master42}
\end{equation}
see eq.~\eqref{IintT}. Here we have defined $J_{ijmn}\equiv\ckm_{ij}\ckm^{-1}_{jm}\ckm_{mn}\ckm^{-1}_{ni}$, for which eq.~\eqref{jarlskog} gives $\im J_{ijmn}=J\epsilon_{im}\epsilon_{jn}$. The notation used for the quark masses should be obvious; here and in the following, the indices $i,j,m,n$ always assume the values $1,2,3$.

To simplify the above expressions, we first use an identity given in eq.~(11.2) of ref.~\cite{salcedo6},
\begin{equation}
\sum_{i,j,m,n}\frac{\epsilon_{im}\epsilon_{jn}}{(\fv p^2+m_{u,i}^2)(\fv p^2+m_{d,j}^2)(\fv p^2+m_{u,m}^2)^2(\fv p^2+m_{d,n}^2)^2}=\frac\Delta{\prod_{q=1}^6(\fv p^2+m_q^2)^2},
\label{epstrick}
\end{equation}
which allows us to rewrite $\hat I^{k,3-k}_{1,1,2,2}$ in the form
\begin{equation}
\hat I^{k,3-k}_{1,1,2,2}=\imag J\Delta\Biggl(\prod_{q=1}^6 \frac{\partial}{\partial m_q^2}\Biggr)\int_{\fv p}\frac{(\vek p^2)^k(p_0^2)^{3-k}}{\prod_{q=1}^6(\fv p^2+m_q^2)}.
\label{I1122}
\end{equation}
Similarly, we obtain for the sum $\hat I^{k,4-k}_{1,1,2,3}+\hat I^{k,4-k}_{1,1,3,2}$ (only this combination appears in the effective action)
\begin{equation}
\begin{split}
\hat I^{k,4-k}_{1,1,2,3}+\hat I^{k,4-k}_{1,1,3,2}=&\imag J\sum_{i,j,m,n}\epsilon_{im}\epsilon_{jn}\int_{\fv p}(\vek p^2)^k(p_0^2)^{4-k}\left(-\frac12\frac{\partial}{\partial\vek p^2}\right)\\
&\times\frac1{(\fv p^2+m_{u,i}^2)(\fv p^2+m_{d,j}^2)(\fv p^2+m_{u,m}^2)^2(\fv p^2+m_{d,n}^2)^2},
\end{split}
\end{equation}
which can be cast in the form
\begin{equation}
\hat I^{k,4-k}_{1,1,2,3}+\hat I^{k,4-k}_{1,1,3,2}=\imag J\Delta\frac{2k+1}4\Biggl(\prod_{q=1}^6 \frac{\partial}{\partial m_q^2}\Biggr)\int_{\fv p}\frac{(\vek p^2)^{k-1}(p_0^2)^{4-k}}{\prod_{q=1}^6(\fv p^2+m_q^2)}=\frac{2k+1}4\hat I^{k-1,4-k}_{1,1,2,2}.
\label{I1123}
\end{equation}
From these expressions, we observe that we only need to evaluate the class of integrals
\begin{equation}
I_{k}\equiv\int_{\fv p}\frac{(\vek p^2)^k(p_0^2)^{3-k}}{\prod_{q=1}^6(\fv p^2+m_q^2)},
\end{equation}
as well as their mass derivatives
\begin{equation}
\tilde{I}_k\equiv \Biggl(\prod_{q=1}^6 \frac{\partial}{\partial m_q^2}\Biggr)I_k,
\label{Itilde}
\end{equation}
for $k=-1,0,1,2,3$.

It would clearly be desirable to evaluate the above integrals in as symmetric a fashion as possible, and preferably in a way, where it is not necessary to specify the number of flavors $N_\text{f}$ ($N_\text{f}=6$ in our case). The calculation is then somewhat simplified by the introduction of a new set of integrals,
\begin{equation}
\begin{split}
H_k^{(0)}&\equiv\int_{\fv p}\frac{(\vek p^2)^k}{\prod_{q}(\fv p^2+m_q^2)},\\
H_k^{(1)}&\equiv\frac{1}{N_\text{f}}\sum_q\int_{\fv p}\frac{(\vek p^2)^k}{\prod_{q'\neq q}(\fv p^2+m_{q'}^2)},\\
H_k^{(2)}&\equiv\frac{1}{N_\text{f}(N_\text{f}-1)}\sum_{\langle q,q'\rangle}\int_{\fv p}\frac{(\vek p^2)^k}{\prod_{q''\notin\{q,q'\}}(\fv p^2+m_{q''}^2)},\\
H^{(n)}_k&\equiv\frac1{N_\text{f}(N_\text{f}-1)\dotsb(N_\text{f}-n+1)}\sum_{\langle q_1,\dotsc,q_n\rangle}\int_{\fv p}\frac{(\vek p^2)^k}{\prod_{q\notin\{q_1,\dotsc,q_n\}}(\fv p^2+m_q^2)},
\end{split}
\label{HinI}
\end{equation}
where the angular brackets indicate that the sum is taken over sets of mutually different indices only. The absence of the frequency variable in the numerator makes these integrals particularly easy to compute. To evaluate them, we next define
\begin{equation}
E_q\equiv\sqrt{\vek p^2+m_q^2},\qquad
\alpha_q\equiv\frac{1}{2E_q}\tanh\frac{E_q}{2T_\text{eff}},\qquad 
\beta_q\equiv \frac1{\prod_{q'\neq q}(m_q^2-m_{q'}^2)},
\end{equation}
as well as the new mass parameters
\begin{equation}
\begin{split}
&M_0\equiv1,\qquad
M_1^2\equiv\frac{1}{N_\text{f}}\sum_q m_q^2,\qquad
M_2^4\equiv\frac{1}{N_\text{f}(N_\text{f}-1)}\sum_{\langle q,q'\rangle}m_q^2 m_{q'}^2,\\
&\cdots,\; \; M_n^{2n}\equiv\frac1{N_\text{f}(N_\text{f}-1)\dotsb(N_\text{f}-n+1)}\sum_{\langle q_1,\dotsc,q_n\rangle}m_{q_1}^2\dotsb m_{q_n}^2.
\end{split}
\end{equation}
Using these quantities, we straightforwardly arrive at the expressions
\begin{equation}
\begin{split}
H_k^{(0)} &=-\int\frac{\dd^3\vek p}{(2\pi)^3} (\vek p^2)^k\sum_q \alpha_q\beta_q,\\
H_k^{(1)}&=\int\frac{\dd^3\vek p}{(2\pi)^3} (\vek p^2)^k\sum_q \alpha_q\beta_q(m_q^2-M_1^2),\\
H_k^{(2)}&=-\int\frac{\dd^3\vek p}{(2\pi)^3} (\vek p^2)^k\sum_q \alpha_q\beta_q(m_q^4-2m_q^2M_1^2+M_2^4),\\
H_k^{(3)}&=\int\frac{\dd^3\vek p}{(2\pi)^3} (\vek p^2)^k\sum_q \alpha_q\beta_q(m_q^6-3m_q^4 M_1^2 +3m_q^2M_2^4-M_3^6),
\end{split}
\end{equation}
which are painless to integrate numerically. Fully analytic expressions can in addition be obtained at zero temperature,
\begin{equation}
\begin{split}
H_k^{(0)}&=\frac{(-1)^{1+k}}{4\pi^{5/2}}\frac{\Gamma(3/2+k)}{\Gamma(2+k)}\sum_qm_q^{2(1+k)}\beta_q\log\frac{m_q}{\Lambda},\\
H_k^{(1)}&=\frac{(-1)^{k}}{4\pi^{5/2}}\frac{\Gamma(3/2+k)}{\Gamma(2+k)}\sum_qm_q^{2(1+k)}(m_q^2-M_1^2)\beta_q\log\frac{m_q}{\Lambda},\\
H_k^{(2)}&=\frac{(-1)^{1+k}}{4\pi^{5/2}}\frac{\Gamma(3/2+k)}{\Gamma(2+k)}\sum_qm_q^{2(1+k)}(m_q^4-2m_q^2M_1^2+M_2^4)\beta_q\log\frac{m_q}{\Lambda},\\
H_k^{(4)}&=\frac{(-1)^{k}}{4\pi^{5/2}}\frac{\Gamma(3/2+k)}{\Gamma(2+k)}\sum_qm_q^{2(1+k)}(m_q^6-3m_q^4 M_1^2 +3m_q^2M_2^4-M_3^6)\beta_q\log\frac{m_q}{\Lambda},
\end{split}
\end{equation}
where the (arbitrary) parameter $\Lambda$ is introduced to make the argument of the logarithm dimensionless but drops from the result upon summation over $q$. As evident from these results, all integrals defined above are both ultraviolet and infrared convergent in four dimensions. 

As the last step of our derivation, we then note that the integrals $I_k$ can be expressed in terms of the $H^{(n)}_k$ via
\begin{align}
\notag I_{-1}=&24M_1^8H_{-1}^{(0)}+6M_2^8H_{-1}^{(0)}-M_4^8H_{-1}^{(0)}+4M_3^6H_{0}^{(0)}+H_{3}^{(0)}+24M_1^6\bigl(H_{0}^{(0)}-H_{-1}^{(1)}\bigr)\\
\notag &-4M_3^6H^{(1)}_{-1}-4H^{(1)}_{2}-12M_1^4\bigl(3M_2^4H^{(0)}_{-1}-H^{(0)}_{1}+2H^{(1)}_{0}-H^{(2)}_{-1}\bigr)\\
\notag &-6M_2^4\bigl(H^{(0)}_{1}-2H^{(1)}_{0}+H^{(2)}_{-1}\bigr)+6H^{(2)}_{1}+4M_1^2\bigl[2M_3^6H^{(0)}_{-1}+H^{(0)}_{2}\\
&-6M_2^4\bigl(H^{(0)}_{0}-H^{(1)}_{-1}\bigr)- 3H^{(1)}_{1}+3H^{(2)}_{0}-H^{(3)}_{-1}\bigr]-4H^{(3)}_{0}+H^{(4)}_{-1},\\
\notag I_{0}=&-6M_1^6H^{(0)}_{0}-M_3^6H^{(0)}_{0}+3M_2^4H^{(0)}_{1}-H^{(0)}_{3}-6M_1^4\bigl(H^{(0)}_{1}-H^{(1)}_{0}\bigr)-3M_2^4H^{(1)}_{0}\\
\notag &+3H^{(1)}_{2}-3M_1^2\bigl(-2M_2^4H^{(0)}_{0}+H^{(0)}_{2}-2H^{(1)}_{1}+H^{(2)}_{0}\bigr)-3H^{(2)}_{1}+H^{(3)}_{0},\\
\notag I_{1} =&2M_1^4H^{(0)}_{1}-M_2^4H^{(0)}_{1}+H^{(0)}_{3}+2M_1^2\bigl(H^{(0)}_{2}-H^{(1)}_{1}\bigr)-2H^{(1)}_{2}+H^{(2)}_{1},\\
\notag I_{2} =&-M_1^2H^{(0)}_{2}-H^{(0)}_{3}+H^{(1)}_{2},\\
\notag I_{3} =&H^{(0)}_{3}.
\end{align}
From here, the desired integrals $\hat I^{k,3-k}_{1,1,2,2}$, $\hat I^{k,4-k}_{1,1,2,3}$ and $\hat I^{k,4-k}_{1,1,3,2}$ are reconstructed using eqs.~\eqref{I1122} and~\eqref{I1123}. Finally, we will for completeness reproduce the explicit expressions of the coefficients $c_i$, appearing in eqs.~\eqref{O0}--\eqref{O2}, in terms of the $\tilde{I}_k$ integrals,
\begin{align}
\notag c_1&=\frac{\tilde{I}_3}{(\cdots)|_{T=0}},\qquad
c_2=\frac{35\tilde{I}_1+70\tilde{I}_2+27\tilde{I}_3}{(\cdots)|_{T=0}},\qquad
c_3=\frac{7\tilde{I}_2+3\tilde{I}_3}{(\cdots)|_{T=0}},\\
\notag c_4&=\frac{35\tilde{I}_1+42\tilde{I}_2+15\tilde{I}_3}{(\cdots)|_{T=0}},\qquad
c_5=\frac{35\tilde{I}_1+42\tilde{I}_2+23\tilde{I}_3}{(\cdots)|_{T=0}},\qquad
c_6=\frac{35\tilde{I}_1+70\tilde{I}_2+19\tilde{I}_3}{(\cdots)|_{T=0}},\\
\notag c_7&=\frac{35\tilde{I}_1+28\tilde{I}_2+9\tilde{I}_3}{(\cdots)|_{T=0}},\qquad
c_8=\frac{3\tilde{I}_0+7\tilde{I}_1+5\tilde{I}_2+\tilde{I}_3}{(\cdots)|_{T=0}},\qquad
c_9=\frac{35\tilde{I}_1+14\tilde{I}_2+11\tilde{I}_3}{(\cdots)|_{T=0}},\\
\label{cs} c_{10}&=\frac{105\tilde{I}_0+105\tilde{I}_1+119\tilde{I}_2-9\tilde{I}_3}{(\cdots)|_{T=0}},\qquad
c_{11}=\frac{105\tilde{I}_1+42\tilde{I}_2+\tilde{I}_3}{(\cdots)|_{T=0}},\\
\notag c_{12}&=\frac{8(7\tilde{I}_2-\tilde{I}_3)}{35(\tilde{I}_0+3\tilde{I}_1+3\tilde{I}_2+\tilde{I}_3)|_{T=0}},\qquad
c_{13}=\frac{4(35\tilde{I}_1+14\tilde{I}_2-5\tilde{I}_3)}{35(\tilde{I}_0+3\tilde{I}_1+3\tilde{I}_2+\tilde{I}_3)|_{T=0}}.
\end{align}
Here, the notation in the first 11 coefficients indicates that they are normalized by the respective zero-temperature results. The normalization of $c_{12},c_{13}$ has been chosen in such a way that the prefactors of the corresponding terms in eq.~\eqref{O1} are equal to one.

\section{Momentum integral in the 6+2 sector}
\label{sec:jarlskog6}

Similarly to eq.~\eqref{master42} of the previous appendix, the momentum integral needed for the evaluation of the CP-violating operators of the 6+2 type at zero temperature reads
\begin{equation}
\begin{split}
\hat I^4_{1,1,1,1,2,2}=\imag\im\sum_{\substack{i,j,k\\ l,m,n}}K_{ijklmn}\int_{\fv p}(\fv p^2)^4&\frac1{(\fv p^2+m_{u,i}^2)(\fv p^2+m_{d,j}^2)(\fv p^2+m_{u,k}^2)(\fv p^2+m_{d,l}^2)}\\
&\times\frac1{(\fv p^2+m_{u,m}^2)^2(\fv p^2+m_{d,n}^2)^2},
\end{split}
\label{I4aux}
\end{equation}
where $K_{ijklmn}\equiv\ckm_{ij}\ckm^{-1}_{jk}\ckm_{kl}\ckm^{-1}_{lm}\ckm_{mn}\ckm^{-1}_{ni}$, analogously to the Jarlskog invariant defined in eq.~\eqref{jarlskog}. In order to evaluate this sum, we first try to simplify the expression for the $K$-tensor.

Suppose first that $i=k$. Thanks to the unitarity of the CKM matrix, one finds $\im K_{ijilmn}=|\ckm_{ij}|^2\im J_{ilmn}=J\epsilon_{im}\epsilon_{ln}|\ckm_{ij}|^2$, thus proving that for $i=k=m$, $K_{ijklmn}$ is real. We conclude from here that its imaginary part can be written as a sum of terms corresponding to three possible cases, in which two of the indices $i,k,m$ are equal, and one case, in which they are all different,
\begin{equation}
\begin{split}
\im K_{ijklmn}=&J\delta_{ki}\epsilon_{km}\epsilon_{ln}|\ckm_{ij}|^2+J\delta_{mk}\epsilon_{mi}\epsilon_{nj}|\ckm_{kl}|^2+J\delta_{im}\epsilon_{ik}\epsilon_{jl}|\ckm_{mn}|^2\\
&+|\epsilon_{ikm}|\im\bigl(\ckm_{ij}\ckm^{-1}_{jk}\ckm_{kl}\ckm^{-1}_{lm}\ckm_{mn}\ckm^{-1}_{ni}\bigr).
\end{split}
\end{equation}
In the last term, we can use the unitarity of the CKM matrix to eliminate the index $m$, $\ckm^{-1}_{lm}\ckm_{mn}=\delta_{ln}-\ckm^{-1}_{li}\ckm_{in}-\ckm^{-1}_{lk}\ckm_{kn}$, leading us to
\begin{equation}
\begin{split}
\frac1J\im K_{ijklmn}=&\delta_{ki}\epsilon_{km}\epsilon_{ln}|\ckm_{ij}|^2+\delta_{mk}\epsilon_{mi}\epsilon_{nj}|\ckm_{kl}|^2+\delta_{im}\epsilon_{ik}\epsilon_{jl}|\ckm_{mn}|^2\\
&+|\epsilon_{ikm}|\epsilon_{ik}\epsilon_{jl}(\delta_{ln}-|\ckm_{in}|^2)-|\epsilon_{ikm}|\epsilon_{ik}\epsilon_{jn}|\ckm_{kl}|^2.
\end{split}
\end{equation}
This form can be further simplified using the trivial identity $|\epsilon_{ikm}|\epsilon_{ik}=\epsilon_{ikm}$ and the fact that $K_{ijklmn}$ is invariant under cyclic permutations of the pairs of indices $ij$, $kl$ and $mn$, allowing us to write the tensor $K$ in a maximally symmetric form
\begin{equation}
\begin{split}
\frac1J\im K_{ijklmn}=&\delta_{ki}\epsilon_{km}\epsilon_{ln}|\ckm_{ij}|^2+\delta_{mk} \epsilon_{mi}\epsilon_{nj} |\ckm_{kl}|^2+\delta_{im}\epsilon_{ik}\epsilon_{jl}|\ckm_{mn}|^2\\
&+\frac13\epsilon_{ikm}\bigl[\epsilon_{ln}(\delta_{nj}+|\ckm_{ij}|^2-|\ckm_{kj}|^2)+\epsilon_{nj}(\delta_{jl}+|\ckm_{kl}|^2-|\ckm_{ml}|^2)\\
&+\epsilon_{jl}(\delta_{ln}+|\ckm_{mn}|^2-|\ckm_{in}|^2)\bigr].
\end{split}
\end{equation}
Plugging this expression back to eq.~\eqref{I4aux} and performing some straightforward albeit tedious manipulations, we arrive at the final expression for the momentum integral in the 6+2 sector,
\begin{equation}
\hat I^4_{1,1,1,1,2,2}=\imag J\Delta\int_{\fv p}\frac{(\fv p^2)^4}{\prod_{q=1}^6(\fv p^2+m_q^2)^2}\sum_{i,j}\frac{|\ckm_{ij}|^2}{(\fv p^2+m_{u,i}^2)(\fv p^2+m_{d,j}^2)},
\label{integral62}
\end{equation}
which is simple to evaluate numerically.


\bibliographystyle{JHEP}
\bibliography{long_paper}

\end{document}